\documentclass[conference]{IEEEtran}
\IEEEoverridecommandlockouts

\makeatletter
\def\thanks#1{\protected@xdef\@thanks{\@thanks
        \protect\footnotetext{#1}}}
\makeatother

\makeatletter
\renewcommand\footnoterule{\relax\kern-5pt
\hrule
\kern4.6pt}
\makeatother

\usepackage{caption}
\usepackage{subcaption}
\usepackage[available,functional,reproduced]{ndssbadges}

\usepackage[normalem]{ulem}

\usepackage[skins]{tcolorbox}
\newtcolorbox{mybox}[2][]{%
  attach boxed title to top center
               = {yshift=-11pt},
  colframe     =black,
  colbacktitle = black,
  title        = #2,#1,
  enhanced,
}
\usepackage{tcolorbox}

\newenvironment{claimproof}[1][\revision{\textit{Proof}}]{%
    \noindent\textit{#1\revision{:}}\quad 
}{%
    \revision{\hfill $\square$} \vspace{10pt} 
}

\usepackage{xspace}
\usepackage{color, colortbl}

\usepackage{enumitem}

\usepackage{graphicx}
\usepackage{adjustbox}

\usepackage{multirow}
\usepackage{multicol}

\usepackage{amssymb}
\usepackage{framed}
\newcommand{\st}{\scriptscriptstyle}

\newcommand{\update}{\ensuremath{\mathtt{Update}}\xspace}

\newcommand{\cl}{\ensuremath{\mathcal{C}}\xspace}
\newcommand{\tp}{\ensuremath{\mathcal{TP}}\xspace}

\newcommand{\simm}{\ensuremath{\mathsf{Sim}}\xspace}

\newcommand{\adv}{\ensuremath{\mathcal{A}}\xspace}

\newcommand{\leak}{\ensuremath{\mathcal{L}}\xspace}
\newcommand{\leakW}{\ensuremath{\mathcal{W}}\xspace}

\newcommand{\gr}{\ensuremath{\mathcal{G}}\xspace}

\newcommand{\gpsi}{\ensuremath{\text{G-PSI}}\xspace}

\newcommand{\pmpd}{\ensuremath{\text{P-MPD}}\xspace}

\newcommand{\fl}{\ensuremath{\text{FL}}\xspace}

\newcommand{\epmpd}{\ensuremath{\text{EP-MPD}}\xspace}

\newcommand{\egpsi}{\ensuremath{\text{EG-PSI}}\xspace}

\newcommand{\egpsione}{\ensuremath{\ensuremath{\text{EG-PSI}^{\st (\text{I})}}}\xspace}
\newcommand{\egpsitwo}{\ensuremath{\text{EG-PSI}^{\st (\text{II})}}\xspace}

\newcommand{\epmpdone}{\ensuremath{\epmpd^{\st (\text{I})}}\xspace}

\newcommand{\epmpdtwo}{\ensuremath{\epmpd^{\st (\text{II})}}\xspace}\xspace

\newcommand{\thetab}{\bm\theta}
\usepackage{bm}

\newcommand{\tee}{\ensuremath{\mathcal{TEE}}\xspace}

\newcommand{\set}{\ensuremath{{S}}\xspace}

\newcommand{\revision}[1]{\textcolor{black}{#1}}

\newcommand{\prf}{\ensuremath{\mathtt{PRF}}}
\usepackage[hidelinks]{hyperref}
\newcommand{\PRP}{\ensuremath{\mathtt{PRP}}\xspace}
\usepackage{colortbl}
\usepackage[skins]{tcolorbox}

\usepackage{mathtools}

\usepackage{amsthm}

\usepackage{amsfonts,amsmath,graphicx,setspace,tipx}

\newtheorem{theorem}{Theorem}

\theoremstyle{definition}
\newtheorem{definition}{Definition}
\newtheorem{claim}{Claim}

\theoremstyle{theorem}
\ifCLASSINFOpdf
\else
\fi
\begin{document}
%



\title{ 
Privacy-Preserving Data Deduplication \\ for Enhancing Federated Learning \\ of Language Models \\ (Extended Version)
}

\author{\IEEEauthorblockN{Aydin Abadi\IEEEauthorrefmark{1}\thanks{\IEEEauthorrefmark{1}Equal contribution. Listing order is alphabetical.}
}
\IEEEauthorblockA{Newcastle University\\
aydin.abadi@ncl.ac.uk}
\and
\IEEEauthorblockN{Vishnu Asutosh Dasu\IEEEauthorrefmark{1}}
\IEEEauthorblockA{Pennsylvania State University\\
vdasu@psu.edu}
\and
\IEEEauthorblockN{Sumanta Sarkar\IEEEauthorrefmark{1}}
\IEEEauthorblockA{University of Warwick\\
sumanta.sarkar@warwick.ac.uk}}

\IEEEoverridecommandlockouts
\makeatletter\def\@IEEEpubidpullup{6.5\baselineskip}\makeatother
\IEEEpubid{\parbox{\columnwidth}{
		Network and Distributed System Security (NDSS) Symposium 2025\\
		24-28 February 2025, San Diego, CA, USA\\
		ISBN 979-8-9894372-8-3\\
		https://dx.doi.org/10.14722/ndss.2025.240641\\
		www.ndss-symposium.org\\
}
\hspace{\columnsep}\makebox[\columnwidth]{}}

\maketitle

\begin{abstract}
Deduplication is a vital preprocessing step that enhances machine learning model performance and saves training time and energy. However, enhancing federated learning through deduplication poses challenges, especially regarding scalability and potential privacy violations if deduplication involves sharing all clients' data. In this paper, we address the problem of deduplication in a federated setup by introducing a pioneering protocol, Efficient Privacy-Preserving Multi-Party Deduplication (\epmpd). It efficiently removes duplicates from multiple clients' datasets without compromising data privacy. \epmpd is constructed in a modular fashion, utilizing two novel variants of the Private Set Intersection protocol. Our extensive experiments demonstrate the significant benefits of deduplication in federated learning of large language models. For instance, 
we observe up to 19.62\% improvement in perplexity and up to 27.95\% reduction in running time while varying the duplication level between 10\% and 30\%. \epmpd effectively balances privacy and performance in federated learning, making it a valuable solution for large-scale applications.


\end{abstract}

\section{Introduction}

%
%

Machine learning (ML) is a data-driven process where
the \textit{quality} of the training data significantly influences the \textit{accuracy} of an ML model. Data generated from real-world applications often lacks organization, leading to issues such as missing values, typos, format mismatches, outliers, or duplicated entries within the raw dataset. To ensure meaningful learning, the collected data must undergo a thorough data cleaning process \cite{ICbook19}. 
Duplicated sequences are prevalent in text datasets.  They can adversely affect the training process of Language Models (LMs). Lee et al. \cite{lee-etal-2022-deduplicating} investigated the Colossal Clean Crawled Corpus (C4) \cite{Raffel-C4} dataset. They discovered a 61-word sequence within C4 that was repeated 61,036 times verbatim in the training dataset and 61 times in the validation set. As a result, the trained model generalized poorly on the dataset, and over $1\%$ of the unprompted model outputs were memorized and copied verbatim from the training dataset. Upon deduplicating the dataset, they reduced memorization by up to $10\times$ and, in certain cases, even improved perplexity by up to $10\%$. 

Additionally, memorization negatively affects the privacy and fairness of LMs \cite{carlini2023quantifyingmemorizationneurallanguage}. Memorization makes language models vulnerable to membership inference attacks \cite{suri2023subjectmembershipinferenceattacks,mattern2023membership}, data extraction attacks \cite{nasr2023scalableextractiontrainingdata,carlini21extracting}, and can lead to copyright violations as LMs can regurgitate training data sequences from copyrighted sources \cite{karamolegkou2023copyrightviolationslargelanguage}. Carlini et al. \cite{carlini2023quantifyingmemorizationneurallanguage} conclude that bigger LMs memorize more and more duplicates increase memorization. Kandpal et al. \cite{kandpal2022deduplicatingtrainingdatamitigates} show that the success of most privacy attacks on LMs is largely due to the duplication in the training datasets. Furthermore, in the \fl setting, malicious clients can exploit the memorization of LMs to extract sensitive information from honest clients' datasets \cite{rashid2024fltrojan}. Therefore, we must adopt data-cleaning practices to improve model utility and reduce the risks of privacy attacks. 

While deduplication can improve model performance and reduce memorization,
it also enhances training efficiency in various aspects \cite{lee-etal-2022-deduplicating}. Removing duplicates reduces GPU time and minimizes training costs in terms of time, money, and carbon footprint. This streamlined process optimizes resource utilization and contributes to more sustainable ML  practices \cite{BGMS21,SGM19,Patterson-etal-carbon19}. Sustainable development is a global priority, and aligning ML practices with this theme is crucial.



Federated learning (\fl) is collaborative learning that allows training across multiple decentralized devices without exchanging data. 
In \fl, devices compute local models based on their data and then share the local model updates with a central server. 
This server aggregates the updates to derive a global model that encapsulates the features of all the local data held by the individual devices \cite{fedml_intro}.
Several real-world applications involve training with \fl such as healthcare \cite{ACKAE22}, smart city \cite{sensor}, and edge computing \cite{XYTWL2021}. Some of these applications pose a risk of duplicates in the local training data across multiple devices. For example, Google Keyboard suggestions from a user's text query rely on \fl \cite{google_keyboard}. Local models are trained in situ on Android phones with user data. In this setting, many text queries typed by the users across multiple phones are the same. 

For an \fl process to be efficient and effective, participating devices must perform deduplication. When a data owner solely intends to remove duplicates from their dataset, the privacy risk is minimal since the data is entirely under the owner's control. In the context of FL, the scenario shifts significantly when deduplication is introduced. Multiple devices participating in FL may possess overlapping data, even after deduplicating their datasets. If they aim to deduplicate the combined dataset, one approach could involve sharing their raw data with each other and checking for intersections. However, this approach compromises privacy.

Hence, there is a necessity for privacy-preserving deduplication in \fl, where participating devices would collaboratively deduplicate the combined datasets in a privacy-preserving fashion. This paper demonstrates how deduplication can be executed in \fl without compromising the privacy of the data belonging to the individual devices. 

\subsection{Overview of Privacy-Preserving Deduplication in \fl}

Consider nodes $D_{\st 1}, \ldots, D_{\st n}$ that are involved in \fl, where each $D_{\st i}$ has a dataset $S_{\st i}$.  
Effectively, the training in \fl is conducted on the union of these datasets, i.e., $\bigcup\limits_{\st i=1}^{\st n} S_{\st i}$. 
If intersections exist among the datasets, \fl will incorporate duplicates, leading to the drawbacks discussed earlier. 
%
Consider nodes $D_{\st 1}$ and $D_{\st 2}$, each with  datasets $S_{\st 1}$ and $S_{\st 2}$, respectively implying that \fl training occurs on $S_{\st 1} \cup S_{\st 2}$. If the intersection $S_{\st 1} \cap S_{\st 2}$ is nonempty, training individually on $S_{\st 1}$ and $S_{\st 2}$ would mean training twice on the duplicate $S_{\st 1} \cap S_{\st 2}$. Thus, one of $D_{\st 1}$ and $D_{\st 2}$ should remove the duplicate $S_{\st 1} \cap S_{\st 2}$. This should be done without harming each other's data privacy. Our solution applies private set intersection (PSI) which securely finds the intersection of $S_{\st 1}$ and $S_{\st 2}$ without revealing any other elements \cite{DBLP:conf/eurocrypt/FreedmanNP04,Adensidora-abadi,AbadiDMT22}. 
Once $D_{\st 1}$ and $D_{\st 2}$ learn $S_{\st 1}\cap S_{\st 2}$ through PSI, $D_{\st 1}$ will train on $S_{\st 1}' = S_{\st 1} \setminus S_{\st 1} \cap S_{\st 2}$ and $D_{\st 2}$ will train on $S_{\st 2}$. According to set theory, it holds that $S_{\st 1}'\cup S_{\st 2} = S_{\st 1} \cup S_{\st 2}$. Hence, the resulting \fl model remains as intended, while the training process is devoid of duplicates, avoiding the associated drawbacks. 
The scenario with two nodes seems straightforward. However, it becomes complicated when more than two nodes participate. 
%
%
We now consider $n$ nodes $D_{\st 1}, \ldots, D_{\st n}$, where each node $D_{\st i}$ has a set $S_{\st i}$,  $\forall i, 1\leq i\leq n$, and  $n > 2$. 
Following the approach used in the two-node case, one might be tempted to (i) apply multi-party PSI to $n$ nodes, (ii)  find their intersection $I_{\st n}$, and (iii) let node $D_{\st 1}$ train on $S_{\st 1}' = S_{\st 1} \setminus I_{\st n}$, and rest of the nodes train on their own dataset $S_{\st i}$. This method removes the duplicates that exist across \textit{all} the nodes. However, this does not detect and remove the duplicates that exist among a subset of nodes. For instance, if there is a subset  of $k$ nodes ($k < n$) with a large intersection $I_{\st k}$, then $I_{\st k}$ will remain in the full training dataset $S_{\st 1}'\bigcup (\bigcup\limits_{\st i = 2}^{\st n} S_{\st i})$ as duplicates. 
A generic multi-party PSI does not help in this case. 
Therefore, we need to consider each pair of nodes and remove the duplicates accordingly.

%

\subsection{Our Contributions}

In this paper, our end goal is to develop a scheme that allows multiple clients to benefit from deduplication while training on their data in a federated learning setup. The first requirement is an efficient deduplication of sets belonging to multiple clients. To address this, we introduce the notion of \emph{Privacy-Preserving Multi-Party Deduplication} (\pmpd) in Section \ref{sec:pmpd}. This essentially describes a functionality that takes input datasets $S_{\st 1}, \ldots, S_{\st m}$ (potentially containing duplicates) from $m$ clients and outputs the datasets $S_{\st 1}', \ldots, S_{\st m}'$ such that $\bigcup\limits_{\st i = 1}^{\st m} S_{\st i}' = \bigcup\limits_{\st i = 1}^{\st m} S_i$, where $S_{\st i}' \cap S_{\st j}' = \emptyset, i \ne j$. We realize that an efficient construction of \pmpd requires a substantially improved PSI protocol that supports scalability. This leads us to introduce a new notion for PSI which we call \emph{Group PSI} (\gpsi) in Section \ref{sec:gpsi}. The functionality of \gpsi takes sets from a group of clients and returns each client the intersection of their sets with all the other clients' sets. We provide constructions of \emph{Efficient Group PSI} (\egpsi) that efficiently realizes \gpsi, namely \egpsione and \egpsitwo in Figures \ref{fig:EG-PSI} and \ref{fig:egpsi-2}, respectively. \egpsione is based on symmetric key primitives such as pseudorandom permutation, while \egpsitwo is developed using an oblivious pseudorandom function, which is a well-known public key primitive. With the building block \egpsi, we build \emph{Efficient Privacy-Preserving Multi-Party Deduplication} (\epmpd) that realizes \pmpd as shown in Figure \ref{fig::construction-PPMD}. We prove the security of \epmpd, \egpsione, and \egpsitwo within the simulation-based paradigm. 
Our construction of \epmpd allows for efficient duplicate removal without compromising clients' data privacy, resulting in an improved model after running federated learning on the deduplicated datasets, as outlined in Figure \ref{fig:fedml-dedup-protocol}.

We perform an extensive experimental evaluation to benchmark \epmpd and the effect of the resulting deduplication on \fl. The overall running time for \epmpdone, which employs \egpsione (symmetric key primitives based), is much less than \epmpdtwo, which utilizes \egpsitwo (public key primitives based). Overall, clients enjoy relatively less computation time in \epmpdtwo. Our protocols can scale to large datasets and client counts. For example, when 50 clients have $2^{\st 19}$ data points in their datasets comprising of $30\%$ duplicates, then \epmpdone takes 1160 seconds and \epmpdtwo takes 7653 seconds; client running time is 641 and 111 seconds in \epmpdone and \epmpdtwo respectively. We experiment with fine-tuning 2 LMs with 7 datasets and 10 clients in \fl. We achieve an improvement of up to 19.62\% in perplexity and up to 27.95\% improvement in GPU training time while varying the duplication level between $10\%$ and $30\%$.



\section{Preliminaries}\label{sec:prelimnaries}

\subsection{Notations and Assumptions}

We define a wrapper function $\update(S, \hat{S})\rightarrow S$ which takes two sets, $S$ and $\hat S$. It updates $S$ by removing from it the elements in set $\hat S$ and returns the updated set $S$. In this paper, by the sum of sets (e.g., $\sum\limits_{\st i = 1}^{\st n} S_{\st i}$) we mean the concatenation of the sets which may result in a multi-set. 
%
%
We denote an empty set by $\emptyset$. We denote a size of vector $\vec{v}$ with $|\vec{v}|$. 
We assume that the server and all the users have access to secure channels among them. 
By the notation, $\mathcal{X} \stackrel{c}{\equiv} \mathcal{Y}$, we mean that the two distributions $\mathcal{X}$ and $\mathcal{Y}$ are computationally indistinguishable.

\label{description-FL}

\subsection{Federated Learning (\fl)}\label{sec::FL}
The concept of \fl was proposed by McMahan et al. \cite{fedml_intro} as a framework for training an ML model where the training data are distributed across multiple devices. In \fl, a server orchestrates the training of a global model by aggregating models locally computed by the clients on their local devices.
Suppose there are $n$ clients, and each client $D_{\st i}$ has a dataset $S_{\st i}$. The server has the initial model $\thetab$. 
It sends $\thetab$ to the clients, and each client performs gradient descent computation on their local dataset $S_{\st i}$ as 
$J_i({S}_{\st i}, \thetab) = \frac{1}{d_{\st i}} \sum\limits_{\st (\textbf{x},y) \in {S}_{\st i}} C(\thetab, (\textbf{x},y) ), $
where $C$ is the cost function and $d_{\st i} = |{S}_{\st i}|$. 
The client $D_{\st i}$ computes the local models as $\thetab_{\st i} \leftarrow \thetab - \eta \nabla J({S}_{\st i}, \thetab),$
where $\eta$ is the learning rate. After receiving the local models from the clients, the server performs an aggregated averaging on the local models to derive the global model as:
\begin{equation}
\Theta = \dfrac{1}{d}\big(d_{\st 1}\thetab_{\st 1} + \ldots + d_{\st n}\thetab_{\st n}\big)
\label{eq:1}
\end{equation}
where $d = \sum\limits_{\st i = 1}^{\st n} d_{\st i}$ is the total size of the dataset ${S} = \sum\limits_{\st i = 1}^{\st n} {S}_{\st i}$, and $\thetab_{\st i}$ is the locally trained model on the dataset ${S}_{\st i}$. This computation is repeated until the model converges.

While this framework achieves a basic level of privacy in which each user's data is not directly sent to the server, it is susceptible to advanced privacy attacks such as membership inference attacks \cite{suri2023subjectmembershipinferenceattacks,mattern2023membership} and data extraction attacks \cite{carlini21extracting, rashid2024fltrojan,fowl2023decepticonscorruptedtransformersbreach,balunovic2022lampextractingtextgradients} as the client models $\theta_{\st i}$ are aggregated in a non-private fashion on the server side. Some privacy-preserving \fl protocols have attempted to mitigate these attacks by securely aggregating the client models. These protocols rely on cryptographic techniques like homomorphic encryption \cite{cheng2021secureboost,provfl}, functional encryption \cite{Xu_2019}, or secure aggregation \cite{bonawitz_new, BIKMMPRS17}.
Additionally, differentially private training techniques \cite{dp_sgd} can be used to ensure differential privacy guarantees on the global model $\Theta$. 
We emphasize that in this paper, our proposed schemes are agnostic to the type of \fl mechanism used.

\subsection{Causal Language Modeling (CLM)}
\label{sec:clm}

Causal Language Modeling (CLM) is a natural language processing task where the goal of the language model is to predict the next word or token given a sequence of tokens. The language model autoregressively generates the next token until a pre-determined sequence length is reached or a special STOP token is generated. Given a sequence of $n$ tokens $\mathbf{Y} = \{y_{\st 1}, y_{\st 2} \ldots, y_{\st n-1}, y_{\st n}\}$, the language model $\Theta$ is trained to learn the following probability distribution:
\begin{equation}
    \label{eq:clm_prob}
    P(\mathbf{Y}) = {\displaystyle \prod_{\st i=1}^{\st n} P(y_{\st i}|y_{\st 1}, \ldots, y_{\st i-1})}
\end{equation}

The CLM training objective is to minimize the negative log-likelihood loss given by:
\begin{equation}
\mathcal{L}(\Theta,\mathbf{Y}) = -\displaystyle \sum_{\st i=1}^{\st n} \log(\Theta(y_{\st i}|y_{\st 1}, \ldots, y_{\st i-1}))
\end{equation}

After training is complete, the text is autoregressively sampled from the language model i.e., $\hat{y}_{\st t < n} \sim \Theta(y_{\st t} | y_{\st 1}, \ldots, y_{\st t-1})$.

The \textit{perplexity} metric is commonly used to evaluate the performance of the language model to determine how well it has learned the probability distribution in Equation~\ref{eq:clm_prob}. The perplexity $PP$ of a sequence $\mathbf{y}$ is defined as:
\begin{equation}
    PP(\mathbf{Y}) = \exp\left(-\frac{1}{n}{\displaystyle \sum_{\st i=1}^{\st n} \log(\Theta(y_{\st i}|y_{\st 1}, \ldots, y_{\st i-1}))}\right)
\end{equation}

A lower perplexity score implies that model $\Theta$ has been trained well to estimate the real-world probability distribution. Informally, a sequence with a low perplexity score implies that the model is less ``surprised'' by a sequence of tokens.

\subsection{Security Model}\label{sec::sec-model}

In this paper, we use the simulation-based paradigm of secure multi-party computation \cite{DBLP:books/cu/Goldreich2004} to define and prove the proposed protocol. Since we focus on the static passive (semi-honest) adversarial model, we will restate the security definition within this context, after outlining the threat model.  

\subsubsection{Threat Model Outline} \revision{
In this paper, three types of parties are involved: clients, a server, and a third party.  
We allow these parties to be corrupted by semi-honest adversaries. In this adversarial model, parties follow the protocols’ instructions. Therefore,  adversaries do not modify the model architecture to better suit their attack or send malicious (global) messages or model parameters. They may try to learn more information (from the messages they exchange) than they are supposed to learn \cite{DBLP:books/cu/Goldreich2004}.
In the protocols, we instantiate the third party using a trusted execution environment (\tee). $\mathcal{TEE}$s are commonly used to instantiate a trusted third party due to their presumed security guarantees. The core assurance about $\mathcal{TEE}$s is that they are fully trusted. However, recent studies have shown that $\mathcal{TEE}$s like Intel SGX are vulnerable \cite{CCXZLL20}. Therefore, in our solution, we do not assume $\mathcal{TEE}$s to be fully trusted and minimize our trust assumption to being semi-honest\footnote{A similar threat model for the semi-honest \tee can be found in \cite{DMM23}.}. Specifically, we assume that \tee does not collude with other parties. Furthermore, we allow any $n-1$ clients to collude with each other (where $n$ is the total number of clients). }

 \subsubsection{Two-party Computation} A two-party protocol $\Gamma$ is captured by specifying a random process that maps a pair of inputs to a pair of outputs (one output for each party). Such process is referred to as a functionality denoted by  $f:\{0,1\}^{\st  *}\times\{0,1\}^{\st  *}\rightarrow\{0,1\}^{\st  *}\times\{0,1\}^{\st  *}$, where $f:=(f_{\st  1},f_{\st  2})$. For every input pair $(x,y)$, the output pair is a random variable $(f_{\st  1} (x,y), f_{\st  2} (x,y))$, such that the party with input $x$ wishes to obtain $f_{\st  1} (x,y)$ while the party with input $y$ wishes to receive $f_{\st  2} (x,y)$. The above functionality can be easily extended to more than two parties.


 \subsubsection{Security in the Presence of Passive Adversaries}  In the passive adversarial model, the party corrupted by such an adversary correctly follows the protocol specification. 
 Loosely speaking, a protocol is secure if whatever can be computed by a corrupt party in the protocol can be computed using its input and output only. 
 In the simulation-based model, it is required that a party’s ``view'' in a protocol's 
 execution can be simulated given only its input and output. This implies that the parties learn nothing from the protocol's execution. 
 
 Formally, in two-party case, party $i$’s view (during the execution of $\Gamma$) on input pair  $(x, y)$ is denoted by $\mathsf{View}_{\st  i}^{\st  \Gamma}(x,y)$ and equals $(w, r^{\st  i}, m_{\st  1}^{\st  i}, ..., m_{\st  t}^{\st  i})$, where $w\in\{x,y\}$ is the input of the $i^{\st  th}$ party, $r_{\st  i}$ is the outcome of this party's internal random coin tosses, and $m_{\st  j}^{\st  i}$ represents the $j^{\st th}$ message this party receives.  The output of the $i^{\st  th}$ party during the execution of $\Gamma$ on $(x, y)$ is denoted by $\mathsf{Output}_{\st  i}^{\st  \Gamma}(x,y)$ and can be generated from its own view of the execution.  
%
\begin{definition}\label{def::simulation-based-def-semi-honest}
Let $f$ be the deterministic functionality defined above. Protocol $\Gamma$ securely computes $f$ in the presence of a static  passive probabilistic polynomial-time (PPT) adversary \adv, if for every \adv in the real model, there exist PPT algorithms $(\mathsf {Sim}_{\st  1}, \mathsf {Sim}_{\st  2})$ such that:
\end{definition}
\vspace{-4mm}
\begin{equation*}
\{\mathsf {Sim}_{\st 1}(x, f_{\st 1}(x,y))\}_{\st x,y}\stackrel{c}{\equiv} \{\mathsf{View}_{\st  1}^{\st \adv, \Gamma}(x,y) \}_{\st x, y}
\end{equation*}
\begin{equation*}
\{\mathsf{Sim}_{\st 2}(y, f_{\st 2}(x,y))\}_{\st x,y}\stackrel{c}{\equiv} \{\mathsf{View}_{\st 2}^{\st \adv, \Gamma}(x,y) \}_{\st x,y}
\end{equation*}


Definition \ref{def::simulation-based-def-semi-honest} can be easily extended to $m>2$ parties.

\subsection{Pseudorandom Function and Permutation}\label{sec:PRP-PRF}
Informally, a pseudorandom function $\prf(.)$ is a deterministic function that takes a key of length $\lambda$ and an input of length $u$; and outputs a value of length $v$ indistinguishable from an output of a truly random function.  More formally, a pseudorandom function can be defined as $\mathtt{PRF}: \{0, 1\}^{\st \lambda}\times \{0,1\}^{\st u} \rightarrow  \{0, 1\}^{\st v}$, where $\lambda$ is the security parameter. 

The definition of a pseudorandom permutation, $\mathtt {PRP}: \{0,1\}^{\st \lambda}\times \{0,1\}^{\st u} \rightarrow   \{0,1\}^{\st u}$, is very similar to that of a pseudorandom function, with a difference; namely, it is required the keyed function $\PRP(k,\cdot)$ to be indistinguishable from a uniform permutation, instead of a uniform function. In cryptographic schemes that involve $\PRP$, sometimes honest parties may be required to compute the inverse of pseudorandom permutation, i.e., $\mathtt{PRP}^{\st -1}(k, \cdot)$, as well. In this case, it would require that $\PRP(k, \cdot)$ be indistinguishable from a uniform permutation even if the distinguisher is additionally given oracle access to the inverse of the permutation.

\subsection{Oblivious Pseudorandom Function}

An Oblivious Pseudorandom Function (OPRF) is a protocol that involves a client and a server. OPRF enables the client with input $x \in \{0,1\}^{\st u}$, and the server with key $k \in \{0,1\}^{\st \lambda}$ to execute an instance of \prf. Informally, the security of an OPRF asserts that, by the completion of OPRF, the client only learns the output of the \prf evaluated on inputs $k$ and $x$, i.e., $\mathtt{PRF}(k,x)$ while the server gains no information, e.g., about the input of the client and the output of \prf. 



\subsection{Trusted Execution Environments}\label{subsec::background_tee}

Trusted Execution Environment ($\tee$), also known as a secure enclave, constitutes a secure processing environment comprising processing, memory, and storage hardware units \cite{PassST17,ZhangZ16}. Within this environment, the code and data residing in them remain isolated from other layers in the software stack, including the operating system. An ideal $\tee$ guarantees the preservation of data integrity and confidentiality. 
%
Given the assumption that the physical CPU remains uncompromised, enclaves are shielded from attackers with physical access to the machine, including the memory and system bus. 
Side-channel attacks on different deployments of $\tee$s have been demonstrated in the literature \cite{TramerZLHJS17}. These attacks pose a threat as they could enable attackers to extract secrets from $\tee$s. Nevertheless, $\tee$s technologies have been evolving to address and mitigate side-channel attacks.

Our security, trust, and system assumptions regarding $\tee$s are conservative. Specifically, as detailed in Sections \ref{sec:Construction-of-pmpd} and \ref{sec::construction-of-gpsi}, our solution (i) avoids disclosing any plaintext messages or private keys to \tee and (ii) does not expect \tee to maintain an extensive storage and memory space or possess strong processing resources. 
Instead, we establish a formal model and construction under the assumption that
\tee guarantees execution integrity (and authenticity), and ensures minimal confidentiality. 
Specifically, we formally demonstrate that \tee at the most only learns the size of the encrypted computation result (i.e., the intersections' cardinality). In our work, \tee can be seamlessly substituted with any semi-honest server that does not collude with other entities.

\section{Group PSI (\gpsi)}
\label{sec:gpsi}
\subsection{Formal Definition  of \gpsi}
In this section, we present the concept of Group PSI (\gpsi). In \gpsi, there are two groups of clients, $\gr_{\st 0}$ and $\gr_{\st 1}$, where each group contains $m$ clients, $\gr_{\st j}:\{\cl_{\st j,1},\ldots,\cl_{\st j,m}\}$, $0\leq j\leq 1$. 
Each client $\cl_{\st j,1}$ has a set $\set_{\st j,1}$, such that no pair of clients' sets in the same group share a common element.

{Informally, \gpsi allows every client in one group to (efficiently) find the intersection that their set has with the set of every client of the other group, without allowing them to learn anything beyond that about other clients' set elements. }
To achieve high computational efficiency in \gpsi, we will involve a third-party \tp that assists the clients with computing the intersection. The functionality ${f}_{\st\gpsi}$ that \gpsi computes takes a set $\set_{\st j, i}$ from every client $\cl_{\st j, i}$ and no input from $\tp$. It returns (i) to every client  $\cl_{\st j, i}$ a vector $\vec{v}_{\st j, i}$ which contains the intersection that $\cl_{\st j, i}$'s set has with every other client's set in the other group and (ii) to \tp an empty set $\emptyset$. Hence, functionality $f_{\st\gpsi}$ can be formally defined as follows. 
\begin{equation}\label{equ::func-with-tp}
\begin{split}
 f_{\st\gpsi} \Big((\underbrace{\set_{\st 0, 1},\ldots, \set_{\st 0, m}}_{\st \gr_{\st 0}}), (\underbrace{\set_{\st 1, 1},\ldots,\set_{\st 1, m}}_{\st \gr_{\st 1}}), \emptyset\Big)\rightarrow\\ \Big((\underbrace{\vec{v}_{\st 0, 1},\ldots, \vec{v}_{\st 0, m}}_{\st \gr_{\st 0}}), (\underbrace{\vec{v}_{\st 1, 1},\ldots,\vec{v}_{\st 1, m}}_{\st \gr_{\st 1}}), \emptyset\Big),
 \end{split}
\end{equation}

where, $\vec{v}_{\st j, i}=\Big[[\set_{\st j, i}\cap \set_{\st 1-j, 1}],\ldots,[\set_{\st j, i}\cap \set_{\st 1-j, m}]\Big]$, $0\leq j \leq 1$, and $1\leq i \leq m$.

In the case where clients of only one of the groups, e.g., $\gr_{\st 0}$, receives the result, then the above functionality simply returns $\emptyset$ to the clients in the other group, e.g., ${\vec{v}_{\st 1, 1}=\ldots=\vec{v}_{\st 1, m}} = \emptyset$.


Since \tp performs computation on all clients' encrypted sets, there is a possibility of leakage to \tp. Depending on the protocol that realizes $f_{\st\gpsi} $ this leakage could contain different types of information; for instance, it could contain (i) the size of the intersection of any two clients' sets, or (ii) the size of the intersection of all clients' sets, or (iii) nothing at all.  Often such leakage is defined by a leakage function $\leak$ that takes all parties (encoded) inputs and returns the amount of leakage.

We assert that a protocol securely realizes $f_{\st\gpsi}$ if (1) it reveals nothing beyond a predefined leakage to \tp and (2) whatever can be computed by a client in the protocol can be obtained from its input and output. This is formalized under the simulation model. We require a client’s view during \gpsi's execution to be simulatable given its input and output. We require that \tp's view can be simulated given the leakage.

\begin{definition}[Security of \gpsi]\label{def::g-psi-def} Let 
$\gr_{\st 0}$ and $\gr_{\st 1}$ be two groups of clients, where each group contains $m$ clients, $\gr_{\st j}:\{\cl_{\st j,1},\ldots,\cl_{\st j,m}\}$, $0\leq j\leq 1$. Let  \leak denote a leakage function, $f_{\st\gpsi}$ be the functionality defined above (in Relation \ref{equ::func-with-tp} on page \pageref{equ::func-with-tp}), $S=\{\set_{\st 0, 1},\ldots, $ $ \set_{\st 0, m}, \set_{\st 1, 1},\ldots,\set_{\st 1, m}\}$, and $\set_{\st j, i}$ represent a set belonging to client $\cl_{\st j, i}$.  Then, a protocol $\Gamma$ securely realizes $f_{\st\gpsi}$ in the presence of a static semi-honest PPT adversary \adv, if for every \adv in the real model, there exists a PPT adversary (simulator) \simm in the ideal model, such that for every  $\cl_{\st j,i}\in\{ \cl_{\st 0,1},\ldots,\cl_{\st 0,m},\cl_{\st 1,1},\ldots,\cl_{\st 1,m}\}$ and $\tp$, Relations \ref{equ::client-sec} and \ref{equ::tee-sec} hold respectively.  
\begin{equation}\label{equ::client-sec}
  \{\mathsf{Sim}_{\st \cl_{\st j,i}}(\set_{\st j,i},\vec{v}_{\st j, i})\}_{\st S}\stackrel{c}{\equiv} \{\mathsf{View}_{\st  \cl_{\st j,i}}^{\st \adv, \Gamma}(S, \emptyset) \}_{\st S}
  \end{equation}
  \begin{equation}\label{equ::tee-sec}
    \{\mathsf{Sim}^{\st \leak}_{\st \tp}(\emptyset, \emptyset)\}_{\st S}\stackrel{c}{\equiv} \{\mathsf{View}_{\st \tp}^{\st \adv, \Gamma}(S, \emptyset) \}_{\st S}
  \end{equation}


where $\vec{v}_{\st j, i}=\Big[[\set_{\st j, i}\cap \set_{\st 1-j, 1}],\ldots,[\set_{\st j, i}\cap \set_{\st 1-j, m}]\Big]$, $0\leq j \leq 1$, $1\leq i \leq m$, and $\tp$'s input is $\emptyset$.


\end{definition}


\subsection{Efficient Construction of \gpsi}\label{sec::construction-of-gpsi}

In this section, we introduce two efficient protocols that realize \gpsi. The first one, called \egpsione, is highly efficient and based on symmetric key cryptography.  The second one, called \egpsitwo, is based on OPRF (in turn depends on public key cryptography) and discloses less information to \tee than the former does. Thus, these two protocols trade-off between performance and leakage amount.


\subsubsection{\egpsione}\label{sec::construction-of-gpsi-1}
%
At a high level, \egpsione operates as follows. Initially, each client in group $\gr_{\st 0}$ agrees with every client in group $\gr_{\st 1}$ on a secret key. Every client encrypts its set elements using every key it has agreed on with other clients (in a different group). 
Each client, for every encrypted set element,
temporally stores a triple that includes (i) the encrypted element, (ii) the key used to encrypt that element, and (iii) the index of the client with whom the key was shared. These triples will enable the client to efficiently (a) retrieve the correct key and (b) identify the client with whom it has the element in common when presented with an encrypted element. Once all elements are encrypted using the corresponding keys, each client transmits only its encrypted elements to \tee. 

Given the sets of encrypted elements from all clients, \tee aggregates these sets and identifies the encrypted elements that appear more than once. Subsequently, \tee forwards to a client those encrypted elements that (1) appear more than once and (2) are among the messages that the client initially sent to \tee. 
Upon receiving each encrypted element from \tee, a client searches its local list of triples to locate the corresponding key and the index $l$ representing a specific client. Utilizing the key, the client decrypts the element and regards the resultant element as one of the elements within the intersection it shares with the $l$-th client.     
For a comprehensive description of \egpsione, refer to Figure \ref{fig:EG-PSI}.

\begin{figure}[!htbp]
\setlength{\fboxsep}{.9pt}
\begin{center}
    \begin{tcolorbox}[enhanced,height=179mm, left=0mm,
    drop fuzzy shadow southwest,
    colframe=black,colback=white]

\begin{itemize}[leftmargin=4.2mm]

\item \textit{Parties}. Trusted execution environment $\tee$, clients in group $\gr_{\st 0}:\{\cl_{\st 0,1},\ldots,\cl_{\st 0,m}\}$, and clients in group $\gr_{\st 1}:\{\cl_{\st 1,1},\ldots,\cl_{\st 1,m}\}$. 

\item \textit{Inputs}. Sets $\set_{\st 0,1}, \ldots, \set_{\st 0, m}, \set_{\st 1,1},\ldots, \set_{\st 1,m}$, where each $\set_{\st j,i}$ belongs to client $\cl_{\st j,i}$, $0\leq j \leq 1$ and $1\leq i \leq m$.

 \item\textit{Outputs}. $\vec{v}_{\st j, i}$ to $\cl_{\st j, i}$, where $\vec{v}_{\st j, i}=\Big[[\set_{\st j, i}\cap \set_{\st 1-j, 1}],\ldots,[\set_{\st j, i}\cap \set_{\st 1-j, m}]\Big]$. 
 
\noindent\rule{\textwidth}{0.1pt}

\end{itemize}
\begin{enumerate}[leftmargin=5.2mm]
\item \underline{\textit{Setup}.}
\begin{enumerate}[leftmargin=5.2mm]
\item\label{EG-PSI-1-setup-key-agr} each client $\cl_{\st 0, i}$ in  $\gr_{\st 0}$ agrees with every client $\cl_{\st 1, l}$ in  $\gr_{\st 1}$ on a secret key $k_{\st i, l}$, by picking a random key $k_{\st i, l}$ and sending it to $\cl_{\st 1, l}$. Client $\cl_{\st 0, i}$ stores this key as $k_{\st i, l}$ while $\cl_{\st 1, l}$ stores this key as $k_{\st l, i}$. 
\item\label{EG-PSI-1-enc-elem} each $\cl_{\st j, i}$ takes the following steps:

\begin{enumerate}

\item encrypts its set elements under keys $k_{\st i,l}$ ($\forall l, 1\leq l \leq m$) as follows,
$\forall e\in \set_{\st j, i}: \PRP(k_{\st i,l}, e)\rightarrow e'_{\st i,l}$. Let set $\set'_{\st j, i}$ contain the encrypted set elements of $\cl_{\st j, i}$ and let set $T_{\st j, i }$ contains all triples of the form  $(e'_{\st i,l}, k_{\st i,l}, l)$.

\item sends $\set'_{\st j, i}$ to \tee and  locally keeps $T_{j, i\st }$. 
\end{enumerate}
\end{enumerate}
\item\label{EG-PSI-1-server-side} \underline{\textit{Finding Encrypted Intersection}.} \tee takes the following steps for each $\cl_{\st j, i}$.
\begin{enumerate}

\item appends to an empty set, $R_{\st j, i}$, every ciphertext that satisfy the following conditions hold:  

\begin{itemize}
    \item it appears more than once in the set $S=\sum\limits_{\st j=0}^{\st 1}\sum\limits_{\st i=1}^{\st m} \set'_{\st j, i}$.
    
    \item it appears in set $\set'_{\st j, i}$.
\end{itemize}

\item sends $R_{\st j,i}$ to $\cl_{\st j,i}$.


\end{enumerate}
\item\label{EG-PSI-1-extract-res} \underline{\textit{Extracting Plaintext Intersection}.} Each $\cl_{\st j, i}$ takes the following steps. 

\begin{enumerate} 
\item constructs a vector $\vec{v}_{\st j,i}=[\vec{v}_{\st j,i, 1},\ldots, \vec{v}_{\st j,i, m}]$, where each vector in $\vec{v}_{\st j,i}$ is initially empty. 

\item\label{step::dec-elem} decrypts  $R_{\st j,i}$'s elements as follows. 
$\forall e'\in R_{\st j,i}:$


\begin{enumerate}
    \item retrieves decryption key $k_{\st i, l}$ and index $l$ from $T_{\st j, i}$ using $e'$.
    \item calls $\PRP^{\st -1}(k_{\st i, l}, e')\rightarrow e$ and appends $e$ to $l$-th vector in $\vec{v}_{\st j,i}$.
    \end{enumerate}

\item considers $\vec{v}_{\st j,i}$ as the result.

\end{enumerate}

\end{enumerate}
\end{tcolorbox}
\end{center}
\caption{First Variant of Efficient Group PSI (\egpsione).}
\label{fig:EG-PSI}
\end{figure}


The only information that \tee learns in \egpsi is the \emph{size} of the intersection of any two clients' sets, called  \emph{pair-wise intersection cardinality}. Below, we formally define it. 

  \begin{definition}[pair-wise intersection cardinality]\label{def::pair-wise-intersection-cardinally} Let $(\underbrace{\set'_{\st 0, 1},\ldots, \set'_{\st 0, m}}_{\st \gr_{\st 0}}), $ $(\underbrace{\set'_{\st 1, 1},\ldots,\set'_{\st 1, m}}_{\st \gr_{\st 1}})$ be two groups $\gr_{\st 0}$ and $\gr_{\st 1}$ of (encrypted) sets. Then, vector $\vec{s}$ represents the pair-wise intersection cardinality: $\vec{s}= [\vec{s}_{\st 0, 1},\ldots,$ $ \vec{s}_{\st 0, m}, \vec{s}_{\st 1, 1},\ldots, \vec{s}_{\st 1, m}]$, where  $\vec{s}_{\st j, i}=\Big[\big|[\set_{\st j, i}\cap \set_{\st 1-j, 1}]\big|,\ldots,\big|[\set_{\st j, i}\cap \set_{\st 1-j, m}]\big|\Big]$, $0\leq j \leq 1$, and $1\leq i \leq m$.

  \end{definition}


\begin{definition}\label{def::leakage} Let $\vec{s}$ be a pair-wise intersection cardinality of two groups $\gr_{\st 0}$ and $\gr_{\st 1}$ of encrypted sets: $(\underbrace{\set'_{\st 0, 1},\ldots, \set'_{\st 0, m}}_{\st \gr_{\st 0}}), (\underbrace{\set'_{\st 1, 1},\ldots,\set'_{\st 1, m}}_{\st \gr_{\st 1}})$, with respect to Definition \ref{def::pair-wise-intersection-cardinally}. Then, leakage function \leak is defined as follows: $\leak\Big((\set'_{\st 0, 1},\ldots, \set'_{\st 0, m}), (\set'_{\st 1, 1},\ldots,\set'_{\st 1, m})\Big)\rightarrow \vec{s}$.

\end{definition}


\begin{theorem}
\label{thm:egpsi-1}
Let $f_{\st\gpsi}$ be the functionality defined in Relation \ref{equ::func-with-tp}. Let \leak be the leakage function presented in Definition \ref{def::leakage}. If $\PRP$ is a secure pseudorandom permutation,  \egpsione (presented in Figure \ref{fig:EG-PSI})  securely realizes $f_{\st\gpsi}$, w.r.t.  Definition \ref{def::g-psi-def}.
\end{theorem}
We refer to Appendix \ref{sec::proof-of-EG-PSI-1} for the proof of Theorem \ref{thm:egpsi-1}.

\subsubsection{\egpsitwo}
In this section, we present \egpsitwo which is the second variant of \egpsi. Note that \tee in \egpsione is able to learn the cardinality of the intersection of each pair of clients' sets.  We aim to reduce this leakage in \egpsitwo. However, \egpsitwo is no longer based on symmetric key cryptography as it depends on OPRF. To the best of our knowledge, all the constructions of OPRFs are based on public key cryptography whose  security depends on some hard problem assumptions. The same holds for \egpsitwo. 


A brief overview of the construction of \egpsitwo is as follows. We have two groups of clients $\gr_{\st 0}$ and $\gr_{\st 1}$ having their own set that they want to find the pairwise intersection. There is a \tee that has a key $k$ for a $\mathtt{PRF}$.  
During the setup, each client interacts with  \tee to encrypt their set using $\mathtt{PRF}(k,\cdot)$ through an OPRF call. Then, each client of group $\gr_{\st 1}$ will send their encrypted set to every client of $\gr_{\st 0}$. Upon receiving an encrypted set from a client of $\gr_{\st 1}$, each client of $\gr_{\st 0}$ determines the intersection with their encrypted set. Once the client finds the intersection, it marks the index of the elements in the intersection and consider the elements with the same index in the unencrypted set as elements in the intersection. Similarly, each client of $\gr_{\st 0}$ shares their encrypted set with each client of $\gr_{\st 1}$ and follows the same steps. Ultimately, every client of $\gr_{\st 0}$ knows the intersection between their set and all the sets belonging to the clients of $\gr_{\st 1}$, and the other way around. We give a detailed description of \egpsitwo in Figure \ref{fig:egpsi-2}.

\begin{figure}[!htbp]
\setlength{\fboxsep}{.9pt}
\begin{center}
    \begin{tcolorbox}[enhanced, left=0mm, height=153mm,
    drop fuzzy shadow southwest,
    colframe=black,colback=white]

\begin{itemize}[leftmargin=4.2mm]

\item \textit{Parties}. Trusted execution environment $\tee$, clients in group $\gr_{\st 0}:\{\cl_{\st 0,1},\ldots,\cl_{\st 0,m}\}$, and clients in group $\gr_{\st 1}:\{\cl_{\st 1,1},\ldots,\cl_{\st 1,m}\}$. 

\item \textit{Inputs}. Sets $\set_{\st 0,1}, \ldots, \set_{\st 0, m}, \set_{\st 1,1},\ldots, \set_{\st 1,m}$, where each $\set_{\st j,i}$ belongs to client $\cl_{\st j, i}$, $0\leq j \leq 1$ and $1\leq i \leq m$.

 \item\textit{Outputs}. $\vec{v}_{\st j, i}$ to $\cl_{\st j, i}$, where $\vec{v}_{\st j, i}=\Big[[\set_{\st j, i}\cap \set_{\st 1-j, 1}],\ldots,[\set_{\st j, i}\cap \set_{\st 1-j, m}]\Big]$. 
 
\noindent\rule{\textwidth}{0.1pt}

\end{itemize}
\begin{enumerate}[leftmargin=5.2mm]

\item \underline{\textit{Setup}.}
\begin{enumerate}[leftmargin=5.2mm]
\item[] \tee generates a secret key $k$ for $\mathtt{PRF}$ that will be used for an OPRF evaluation.

\end{enumerate}

\item\label{EG-PSI-2-enc-set} \underline{\textit{Encryption of each client's set}.}
\begin{enumerate}[leftmargin=5.2mm]
\item[] Each client $\cl_{j, i}$  and \tee run an OPRF protocol and obtains the encrypted set $\mathtt{PRF}(k,\set_{\st j, i})$, where $0\leq j \leq 1$ and $1 \leq i \leq m$.

\end{enumerate}

\item\label{EG-PSI-2-find-intersection} \underline{\textit{Finding Encrypted Intersections}.} 
\begin{enumerate}

\item Each client $\cl_{\st j, i}$ sends each $\mathtt{PRF}(k,\set_{\st j, i})$ to every client $\cl_{\st 1-j, i}$ (in the other group), where $0\leq j \leq 1$ and $1 \leq i \leq m$.




\item Each client $\cl_{\st j, i}$, takes the following steps.

\begin{enumerate}
\item Creates an empty set $R_{j,i,t}$ for all $1 \leq t \leq m$.


\item Let $e_\ell$ represents an element of $\mathtt{PRF}(k,\set_{\st j, i})$, where $1 \leq \ell \leq |\mathtt{PRF}(k,\set_{\st j, i})|$. \\
Performs $R_{j,i,t} \leftarrow R_{j,i,t} \cup \{\ell\}$ if $e_\ell$ appears both in $\mathtt{PRF}(k,\set_{\st j, i})$ and $\mathtt{PRF}(k,\set_{\st j-1, i})$.


\end{enumerate}

\end{enumerate}

\item\label{EG-PSI-2-extract} \underline{\textit{Extracting Plaintext Intersection}.} 

\item[] Each $\cl_{j, i}$ for $j = 0, 1$, and $1 \leq i \leq m$ does the following. 

\begin{enumerate}

\item Constructs a vector $\vec{v}_{\st j,i}=[\vec{v}_{\st j,i, 1},\ldots, \vec{v}_{\st j,i, m}]$, where each vector in $\vec{v}_{\st j,i}$ is initially empty.

\item Appends the $\ell$-th element of $\set_{\st j, i}$ to $\vec{v}_{\st j,i, t}$ for all $\ell \in R_{j,i,t}$.

\item Returns $\vec{v}_{\st j,i}$.

\end{enumerate}

\end{enumerate}
\end{tcolorbox}
\end{center}
\caption{Second Variant of Efficient Group PSI (\egpsitwo).} 
\label{fig:egpsi-2}
\end{figure}

In \egpsitwo, \tee does not participate in the protocol except the OPRF evaluation and that is also a one-time activity. As a result, \tee does not learn anything about the set intersection, it only learns the cardinality of each client's set. 









\begin{theorem}
\label{thm:egpsi-2}
Let $f_{\st\gpsi}$ be the functionality defined in Relation \ref{equ::func-with-tp} (on page \pageref{equ::func-with-tp}). Also, let \leak be the leakage function presented in Definition \ref{def::leakage} with the empty output. If OPRF is secure, then \egpsitwo  securely realizes $f_{\st\gpsi}$, w.r.t. Definition \ref{def::g-psi-def}.
\end{theorem}

Appendix \ref{sec::proof-of-EG-PSI-2} outlines the proof of Theorem \ref{thm:egpsi-2}. 
Note that both variants remain secure in the presence of semi-honest adversaries. However, if \tee or a client is a malicious (or active) adversary, then it can affect the correctness of the result, without being detected.



\section{Privacy-Preserving Multi-Party Deduplication (\pmpd)}
\label{sec:pmpd}
\subsection{Formal Definition of \pmpd}

\pmpd considers the setting where there are $n$ clients $C=\{
\cl_{\st 1},\ldots, \cl_{\st m}\}$ and each $\cl_{\st i}\in C$ has a set $S_{\st i}$. \pmpd enables the clients to (efficiently) remove duplicates from their local sets such that after the deduplication the concatenation of all clients' local sets equals the union of their initial sets.

The functionality $f_{\st \pmpd}$ that \pmpd computes takes a set $S_{\st i}$ from each $\cl_{\st i}$. It returns an updated (i.e., deduplicated) set $S'_{\st i}$ to each $\cl_{\st i}$. Formerly, $f_{\st \pmpd}$ can be defined as follows:
\begin{equation}\label{relation::pmpd-func}
   f_{\st \pmpd}(S_{\st 1},\ldots,S_{\st m})\rightarrow (S'_{\st 1},\ldots,S'_{\st m}),
\end{equation}
 where $\sum\limits^{\st m}_{\st i=1}S'_{\st i}=\bigcup\limits_{\st i=1}^{\st m}S_{\st i}=S_{\st \cup}$ and $S_{\cup}$ denotes the union of all initial sets and contains only unique elements.

To maintain generality, we define a leakage function, denoted as $\leakW$. The output of this function relies on the protocol implementing $f_{\st \pmpd}$. We assert that a protocol securely realizes $f_{\st \pmpd}$ if whatever can be computed by a party in the protocol can be derived solely from its individual input and output, given the output of $\leakW$. Below, we formally state it.

\begin{definition}[Security of \pmpd]\label{def:security-of-pmpd}

Let $S=\{\set_{\st 1},\ldots, \set_{\st  m}\}$, and $\set_{\st i}$ represent a set belonging to client $\cl_{\st i}$. Let \leakW denote a leakage function and  $f_{\st \pmpd}$ be the functionality defined in Relation \ref{relation::pmpd-func}. Then, a protocol $\Psi$ securely realizes $f_{\st \pmpd}$ in the presence of a static semi-honest PPT adversary \adv, if for every \adv in the real model, there exists a PPT \simm in the ideal model, such that for every  $\cl_{\st i}\in\{ \cl_{\st 1},\ldots,\cl_{\st m}\}$, Relation \ref{equ::pmpd-client-sec} holds. 
  \begin{equation}\label{equ::pmpd-client-sec}
  \{\mathsf{Sim}^{\st \leakW}_{\st \cl_{\st i}}(\set_{\st i}, \set'_{\st i})\}_{\st S}\stackrel{c}{\equiv} \{\mathsf{View}_{\st \cl_{\st i}}^{\st \adv, \Psi}(S) \}_{\st S}.
  \end{equation}
\end{definition}

\subsection{Constructions of \pmpd}

\subsubsection{Construction Based on \egpsi}
\label{sec:Construction-of-pmpd}
We introduce a pioneering protocol called Efficient Privacy-Preserving Multi-Party Deduplication (\epmpd) that realizes \pmpd efficiently. The idea behind  \epmpd's design involves constructing a binary tree with leaf nodes representing the clients’ indices. At each level, each cluster is formed with two different groups of clients, namely $\gr_{\st 0}$ and $\gr_{\st 1}$. Subsequently,  \egpsi is recursively applied to the sets of clients sharing the same cluster until the tree's root is reached. After each invocation of \egpsi, the clients of {$\gr_{\st 0}$} update their sets by removing the intersections, returned by \egpsi, from their local sets. These updated sets are then used as input in the subsequent \egpsi invocation. 

\begin{figure}[ht]
    \centering
    \begin{subfigure}{0.47\textwidth}
        \centering
        \includegraphics[width=\linewidth, height=4cm]{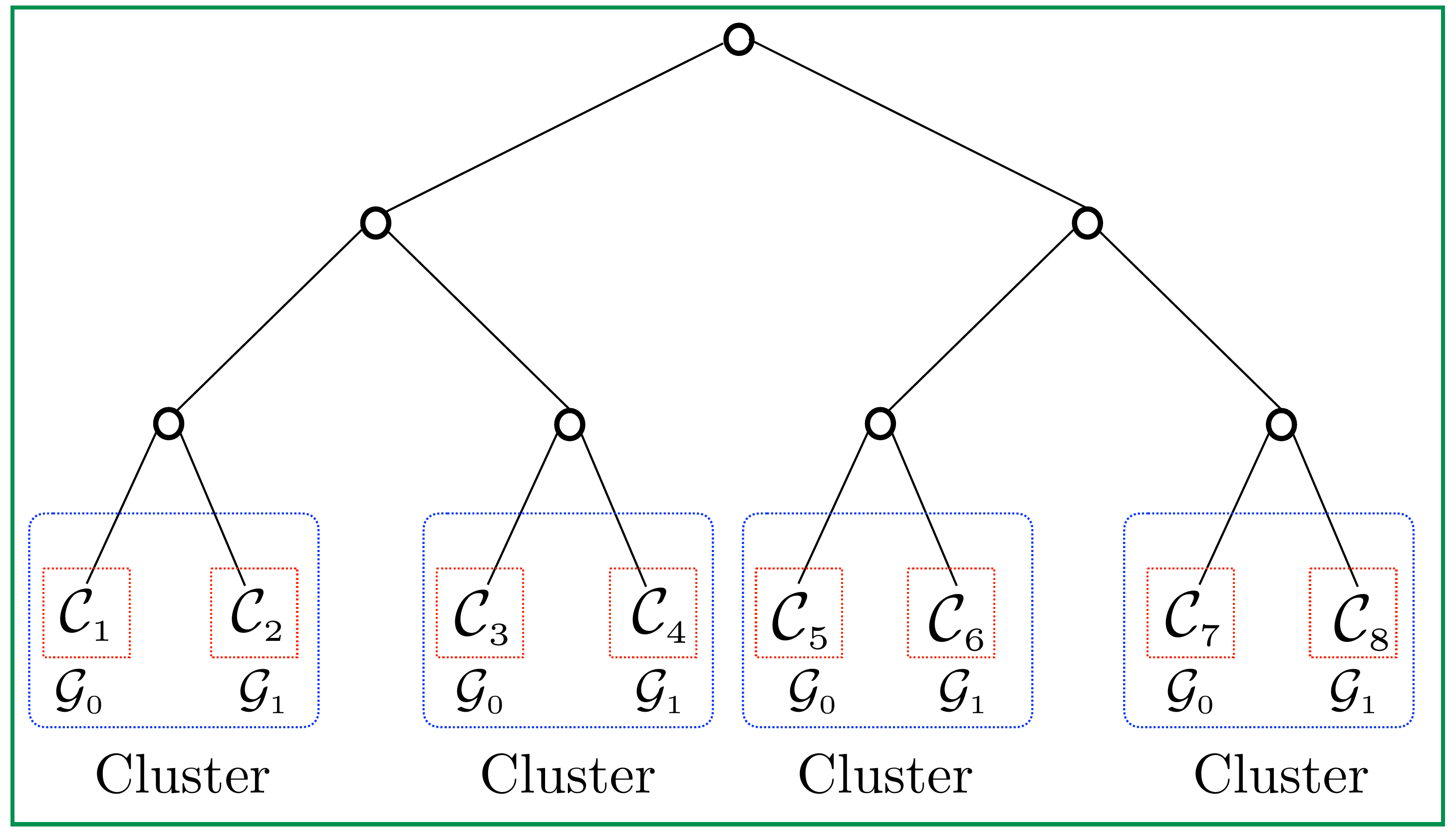}
        \caption{Clients joining various groups and clusters at level 1.}\label{fig:Diag-1}
    \end{subfigure}
%
%
    \begin{subfigure}{0.47\textwidth}
        \centering
        \includegraphics[width=\linewidth, height=4cm]{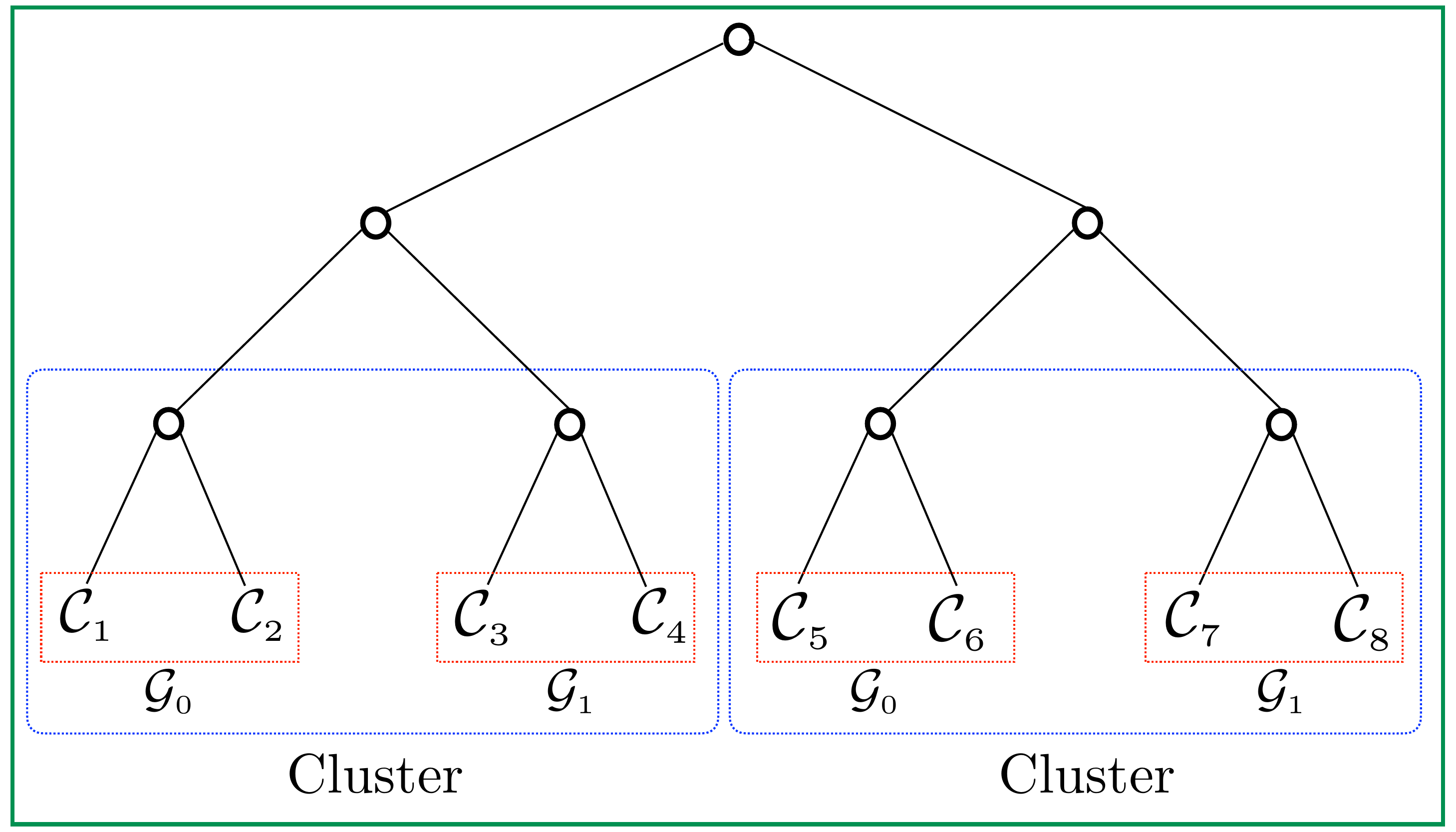}
        \caption{Clients joining various groups and clusters at  level 2.}\label{fig:Diag-2}
    \end{subfigure}
    \hfill
    \begin{subfigure}{0.47\textwidth}
        \centering
        \includegraphics[width=\linewidth, height=4cm]{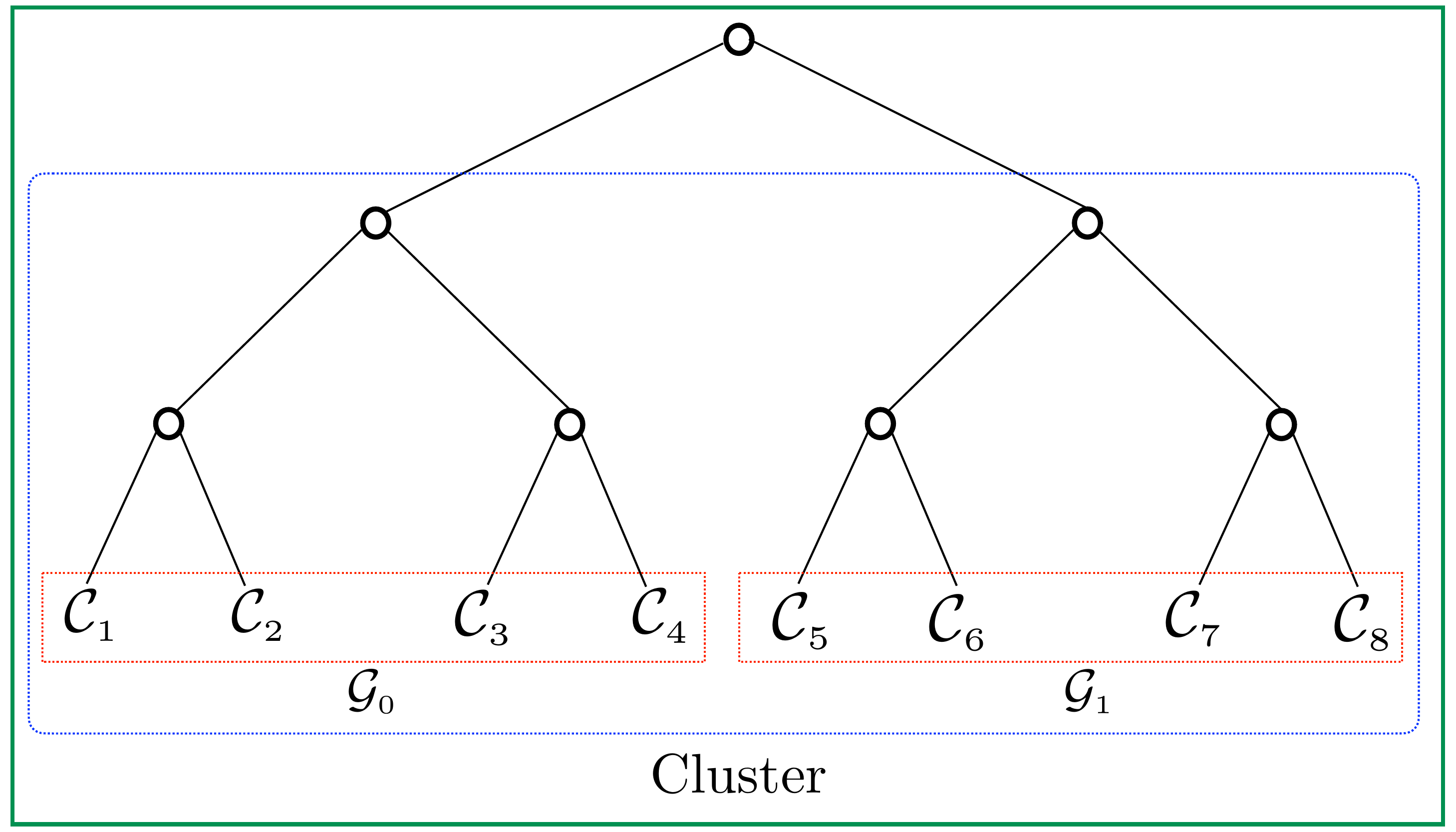}
        \caption{Depiction of how clients become part of diverse groups and clusters at  level 3.}\label{fig:Diag-3}
    \end{subfigure}

    \caption{The process of eight clients forming clusters and groups at various levels of a binary tree. At each level, clients belonging to groups $\gr_{\st 0}$ and $\gr_{\st 1}$ within the same cluster initiate \egpsi, followed by the updating of their respective local sets.}\label{fig::clients-formation}
\end{figure}

Next, we delve into further detail. We begin by sorting the clients' indices in ascending order. Next, we construct a binary tree such that the leaf nodes represent the clients' indices.

At level $d=1$, commencing from the left-hand side, every two clients form a cluster. Within each cluster, the first client is assigned to group $\gr_{\st 0}$ and the second client to group $\gr_{\st 1}$. 
Then, the clients within the same cluster engage in an instance of \egpsi. Upon the return of the sets' intersection by \egpsi, only the client situated on the left-hand side, specifically the one with the smaller index, modifies its local set by eliminating the elements of the intersection from its sets. This process is outlined in Figure \ref{fig:Diag-1}.


At level $d=2$, starting from the left-hand side, every set of $2^{\st d}$ clients form a cluster. Within each cluster, the initial $\frac{2^{\st d}}{2}=2$ clients enter group $\gr_{\st 0}$ and the remainder (i.e., the remaining $2$ clients) enter group $\gr_{\st 1}$.  Subsequently, the clients within the same cluster engage in an instance of \egpsi. Once again, upon the return of their sets' intersection by \egpsi, the clients possessing the smaller indices update their local sets. This procedure is illustrated in Figure \ref{fig:Diag-2}.
%
The process continues until level $d=\log_{\st 2}{m}$ is reached. At this point, all $m$ clients collectively engage in an instance of \egpsi. Each $\frac{2^{\st d}}{2}=\frac{m}{2}$ clients enter group $\gr_{\st 0}$ and the remaining $\frac{m}{2}$ clients enter group $\gr_{\st 1}$.  These $m$ clients jointly participate in an instance of \egpsi. Similarly to previous steps, the clients with the smaller indices update their local sets based on the output of \egpsi. This procedure is outlined in Figure \ref{fig:Diag-3}.

\begin{figure}[!htbp]
\setlength{\fboxsep}{.9pt}
\begin{center}
    \begin{tcolorbox}[enhanced, left=0mm,height=128mm, 
    drop fuzzy shadow southwest,
    colframe=black,colback=white]

\begin{itemize}[leftmargin=4.2mm]

\item \textit{Parties}. A set of clients $\{\cl_{\st 1},\ldots,\cl_{\st m}\}$. 

\item \textit{Inputs}. Sets $\set_{\st 1}, \ldots, \set_{\st  m}$, where each $\set_{\st i}$ belongs to client $\cl_{\st i}$, $1\leq i \leq m$, and $m$ is a power of 2. 

 \item\textit{Outputs}. Updated sets $\set'_{\st 1}, \ldots, \set'_{\st  m}$, where each $\set'_{\st i}$ belongs to client $\cl_{\st i}$, and $\sum\limits^{\st m}_{\st i=1}S'_{\st i}=\bigcup\limits_{\st i=1}^{\st m}S_{\st i}$.

\end{itemize}
\noindent\rule{\textwidth}{0.1pt}

The clients take the following steps $\forall d$, $1\leq d \leq \log_{\st 2}(m):$ 
\begin{enumerate}[leftmargin=5.2mm]

    \item\underline{\textit{Forming Clusters.}} Every distinct $2^{\st d}$ clients enter the same cluster. Thus, for every level $d$, there will be $\frac{m}{2^{\st d}}$ cluster(s).
    For instance, when $d=1$, then there are the following clusters: $(\cl_{\st 1}, \cl_{\st 2}),\ldots,(\cl_{\st m-1}, \cl_{\st m})$.

    
    \item\underline{\textit{Forming Group.}} In each cluster, the first $\frac{2^{\st d}}{2}$ clients enter group $0$, $\gr_{\st 0}$, and the rest of the clients enter group $1$, $\gr_{\st 1}$. 

    \item\label{DE-find-intersection}\underline{\textit{Finding Intersection.}} In each cluster, the clients of $\gr_{\st 0}$ and  $\gr_{\st 1}$ together engage an instance of \egpsi. Each client $\cl_{\st j}$ receives a vector $\vec{v}_{\st j}$ from \egpsi, where $\vec{v}_{\st j}$ specifies the elements that  $\cl_{\st j}$ has in common with each client of the other group in the same cluster. 
    It would be sufficient if only clients of $\gr_{\st 0}$ in each cluster were aware of the intersection result.
    
    \item\label{DE-remove-duplicates}\underline{\textit{Removing Duplication.}} For every element $e\in \vec{v}_{\st j}$, each $\cl_{\st j}$ in $\gr_{\st 0}$ 
    %
    calls $\update(S_{\st j}, \{e\})\rightarrow S_{\st j}$, where $S_{\st j} \leftarrow S_{\st j} \setminus \{e\}$.

\item  \revision{When $d=\log_{\st 2}({m})$, 
 %
 as all clients have reached the root and invoked EG-PSI at each level, they can ensure that all duplicates have been removed (due to the correctness of our EG-PSI).}

\end{enumerate}

\end{tcolorbox}
\end{center}
\caption{Efficient Privacy-Preserving Multi-Party
Deduplication (\epmpd): the construction of \pmpd.} 
\label{fig::construction-PPMD}
\end{figure}

Figure \ref{fig::construction-PPMD} presents a detailed description of \epmpd. For the sake of simplicity, we assume that $m$ is a power of 2 in \epmpd's description. However, we do not impose on the number of clients. Non-powers of 2 will result in incomplete sub-trees in the formation of clusters in the DE protocol where each group may have an unequal number of clients. This does not affect the \epmpd protocol execution.

\subsection{Naive Approaches}

\subsubsection{Running a Two-Party PSI Multiple Times}\label{naive_approach} One might be tempted to allow each client to invoke an existing two-party PSI with every other client. However, this approach in total requires 
${m\choose 2} = O(m^{\st 2})$ invocations of the two-party PSI. 
Whereas \epmpd requires only $ \sum\limits^{\st \log_{\st 2}(m)}_{\st i=1}(\frac{m}{2^{\st i}})=m-1$ invocations of our efficient tailor-made PSI, \egpsi. To provide concrete values, when $m=256$, the naive scheme requires 32,640 invocations of a standard two-party PSI, whereas our scheme requires only $255$ invocations of \egpsi. 

\subsubsection{Running Multi-Party PSI Once}

One might be tempted to execute an existing multi-party PSI protocol only once on all clients' sets, hoping to identify and subsequently remove duplicated elements.  
Nevertheless, this approach falls short of identifying all duplications. The limitation arises from the fact that a multi-party PSI can uncover elements common to \textit{all} clients, making it incapable of detecting elements shared among only a subset of clients.

\subsubsection{Deduplicating all Sets at Once} 

There is a simplified version of \epmpd. In this variant, each client reaches an agreement with the others on a secret key. Subsequently, each client encrypts its set elements using every key agreed upon with the other clients. Finally, all clients transmit their encrypted set elements to \tee. 
Upon receiving the encrypted sets from all clients,  \tee identifies all duplicated elements. It then transmits to each client: (1) the duplicated elements found in the original elements that the respective client sent to \tee, and (2) the index of the client with which it shares the same element. Based on \tee's response, the clients with smaller indices update their local sets accordingly.

However, this approach requires that \tee maintains a local storage or memory space equal to the size of the concatenation of all encrypted sets. In contrast, \epmpd with the underlying \egpsi building block, allows \tee to have a substantially smaller storage or memory space. This is achieved because, before \tee receives all clients' encrypted sets at level $d=log_{\st 2}(m)$, clients apply updates to their sets multiple times. This iterative process progressively reduces the size of the clients' sets, and the reduction can be particularly significant when the number of duplications is substantial.

 \subsection{Security Analysis}
In any secure (multi-party) deduplication protocol, a participating client will learn the exact redundant elements within its set upon the protocol's completion. 
However, in \epmpd, a client gains slightly more information. To be specific, upon the completion of \epmpd, each client acquires the knowledge of the index (or identity) of the client that shares the redundant element. 
This additional information gained by a client during the execution of \epmpd is formally defined as the output of a leakage function \leakW which we define below.

\leakW takes as input (i) sets $\set_{\st 1}, \ldots, \set_{\st m}$, where each set $\set_{\st i}$ belongs to client  $\cl_{\st i}$, and (ii) sets $R_{\st 1}, \ldots, R_{\st m}$, where each $R_{\st i}$ contains the redundancy (or intersection) that a set $\set_{\st i}$ has with every other set.  \leakW  outputs to the $i$-th client a set $Q_{\st i}$ of triples, where each triple is of the form $(r_{\st i, l}, i, l)$, $r_{\st i, l}
\in R_{\st i}$ is a redundancy between sets $\set_{\st i}$ and $\set_{\st l}$.

\begin{definition}\label{def::leakage-W}  
Let $S=\{\set_{\st 1}, \ldots, \set_{\st m}\}$ be a set of sets, where each set $\set_{\st i}$ belongs to client  $\cl_{\st i}$. Let $R=\{R_{\st 1}, \ldots, R_{\st m}\}$ be a set of sets, where each set $R_{\st i}$ comprises the redundancy that  $\set_{\st i}$ has with every other $j$-th set, for all $j$, $1 \leq j \leq m$ and $ i\neq j$. Then, leakage function \leakW is defined as: $\leakW (S, R)\rightarrow(Q_{\st 1},\ldots, Q_{\st m})$, where each $Q_{\st i}$ is a set of triples of the form $(r_{\st i, l}, i, l)$, $r_{\st i, l}
\in R_{\st i}$ is a redundancy between sets $\set_{\st i}$ and $\set_{\st l}$.
\end{definition}

\begin{theorem}\label{theorem::securit-of-DE}
Let $f_{\st \pmpd}$ be the functionality defined in Relation \ref{relation::pmpd-func} and \leakW be the leakage function presented in Definition \ref{def::leakage-W}. If \egpsi is secure (w.r.t. Definition \ref{def::g-psi-def}), then \epmpd (presented in Figure \ref{fig::construction-PPMD}) securely realizes $f_{\st \pmpd}$, w.r.t. Definition \ref{def:security-of-pmpd}. 
\end{theorem}


The proof of Theorem \ref{theorem::securit-of-DE} can be found in Appendix \ref{sec::proof-of-DE}.

\section{Enhanced \fl: Applying  \epmpd to \fl}


With all essential building blocks in place, we are prepared to elucidate the complete functionality of the system.
There are clients $\cl_{\st 1},\ldots,\cl_{\st m}$ having the sets $\set_{\st 1}, \ldots, \set_{\st  m}$, where each $\set_{\st i}$ belongs to client $\cl_{\st i}$ respectively.
Phase \ref{Local-Deduplication} involves each client conducting local deduplication. Since the local deduplication is a private computation there is no privacy requirement. The data owners can apply deduplication techniques as illustrated in \cite{lee-etal-2022-deduplicating} which results in having a set ${S}_{\st i}'$ free of duplicates by each client $\cl_{\st i}$. 
Subsequently, in Phase \ref{Global-Deduplication} they employ \epmpd to eliminate duplicates across all clients. At the end of this phase, client $\cl_{\st i}$ gets an updated set ${S}_{\st i}''$, such that $\sum\limits_{i = 1}^{\st m} {S}_{\st i}'' =\bigcup\limits_{i = 1}^{\st m}   {S}_{\st i}'.$
Moving to Phase \ref{Federated-learning-phase}, they adopt an \fl protocol (outlined in Section \ref{sec::FL}) to train the model. Further details of this procedure are provided in \autoref{fig:fedml-dedup-protocol}{}. \revision{Appendix \ref{sec::Proof-of-Correctness}, proves the scheme's correctness.}

 \begin{figure}[!htbp]
\setlength{\fboxsep}{.9pt}
\begin{center}
    \begin{tcolorbox}[enhanced, height=130mm, left=0mm,
    drop fuzzy shadow southwest,
    colframe=black,colback=white]

\begin{itemize}[leftmargin=4.2mm]

\item \textit{Parties}. A set of clients $\{\cl_{\st 1},\ldots,\cl_{\st m}\}$. 

\item \textit{Server}. Holds the initial model $\thetab$. 

\item \textit{Inputs}. Sets $\set_{\st 1}, \ldots, \set_{\st  m}$, where each $\set_{\st i}$ belongs to client $\cl_{\st i}$, $1\leq i \leq m$, and $m$ is a power of two. 

 \item\textit{Outputs}. A global updated model $\Theta$.

\end{itemize}
\noindent\rule{\textwidth}{0.1pt}

\begin{enumerate}[leftmargin=5.2mm]

    \item\label{Local-Deduplication}\underline{\textit{Local Deduplication.}} Each client runs a deduplication algorithm on their local dataset. At the end, client $\cl_{\st i}$ receives an updated dataset ${S}_{\st i}'$.

    
    \item\label{Global-Deduplication}\underline{\textit{Global Deduplication.}} \begin{enumerate}
			
				\item All the clients participate in the \epmpd as described in \autoref{fig::construction-PPMD}.
                \item Each client $\cl_{\st i}$ gets updated set ${S}_{\st i}''$, such that $$\sum\limits_{i = 1}^{\st m} {S}_{\st i}'' =\bigcup\limits_{i = 1}^{\st m}   {S}_{\st i}'.$$
                \end{enumerate}

    \item\label{Federated-learning-phase}\underline{\textit{Federated learning.}} 
			\begin{enumerate}
                \item The server and clients agree upon an \fl protocol for training.
				\item The server initiates the learning by sharing the initial model $\thetab$ with each client.
                \item\label{step::local-training-client} Each client $\cl_{\st i}$
                trains on their local dataset ${S}_{\st i}''$ and updates $\thetab$ to $\thetab_{\st i}$.
                
                \item The clients and server aggregate the local models $\thetab_{\st i}$ trained by the clients.
                \item\label{step::server-gen-model} The server outputs the global updated model $\Theta$ for the next training round. 
                \item \revision{Server and clients repeat the above steps \ref{step::local-training-client}--\ref{step::server-gen-model}, until the convergence is reached.}
			\end{enumerate}	       

\end{enumerate}

\end{tcolorbox}
\end{center}
\caption{Federated learning with deduplication.} 
\label{fig:fedml-dedup-protocol}
\end{figure}

\section{Experiments}
\subsection{Implementation Details}

We implement the \epmpd protocol in \texttt{python=3.10}. We use AES with a 128-bit key in CBC mode from the \texttt{cryptography} library as a PRP in \egpsione. To realize an OPRF in \egpsitwo, we use the OPRF algorithm proposed in  \cite{burns_oprf} from the \texttt{oprf} library. All \epmpd experiments are conducted on a batch computing facility with a Dual Intel Xeon E5-2660 v3 @ 2.6 GHz CPU and 60 GB RAM.

For the \fl experiments, we implement an \fl setting for CLMs using \texttt{python=3.10} with the \texttt{transformers=4.40.1} and \texttt{torch=2.3} libraries. We implement the FedAvg \cite{fedml_intro} algorithm for \fl training to analyze the effect of \epmpd on \fl's performance. However, in practice, we recommend combining \epmpd with privacy-preserving \fl protocols for end-to-end privacy guarantees. All \fl experiments are conducted on a server with an Intel(R) Xeon(R) Gold 6336Y CPU @ 2.40GHz CPU and NVIDIA RTX A6000 GPU. \revision{Appendix~\ref{sec:additional_experiments} contains additional experiments with large number of clients and experiments that simulate real-world network delay and bandwidth.}

We run all clients sequentially on a machine for both experiments. However, to emulate a real-world setting where clients run on different devices at the same time, we first compute the individual client running times and then estimate the real-world distributed running time by assuming the clients have run simultaneously. 
We open-source our implementations\footnote{\url{https://github.com/vdasu/deduplication}}.


While implementing \epmpd, we have two options for the \egpsi module: \egpsione and \egpsitwo. We observe that \egpsitwo-based \epmpd calls an OPRF function for every client in the leaf nodes of the \epmpd tree, returning an encryption of the clients' sets. As clients move to upper levels, their set elements remain unchanged, with only the cardinality potentially changing. Therefore, except at the bottom level, OPRF calls are unnecessary for encryption, allowing the same encrypted set or their subsets to be reused at other levels. 
Thus, we have slightly adjusted the implementation.
This significantly reduces time without compromising security properties, as reflected in our experiments. 

\subsection{\epmpd Benchmarks}
\label{sec:benchmarks}

\subsubsection{Experimental Setup}

We designed three experiment settings to benchmark the performance of \epmpd. The three settings analyze the effect of the number of clients, dataset size, and duplication percentage on the runtime of \epmpd. All experiments were performed with both \egpsione and \egpsitwo as building blocks. The duplication percentage refers to the percentage of samples in a client's dataset present in another client's dataset. The three experiment settings are:

\begin{enumerate}[label=\alph*.]

\item The dataset size $|S_{\st i}|$ is fixed at $2^{\st 19}$, the duplication percentage at $30\%$, while the number of clients $m$ is varied. Specifically, $m$ is set to each value in  $\{10, 20, 30, 40, 50\}$.

\item The number of clients is fixed at $50$ and the duplication percentage at $30\%$, while the dataset size $|S_{\st i}|$ is varied. Specifically, $|S_{\st i}|$ is set to each value in $\{2^{\st 10}, 2^{\st 11}, 2^{\st 13}, 2^{\st 15}, 2^{\st 17}, 2^{\st 19}\}$.

\item The dataset size $|S_{\st i}|$ is fixed at $2^{\st 19}$, the number of clients $m$ at $50$, while the duplication percentage is varied. Specifically, it is set to each value in $\{10\%, 30\%, 50\%, 70\%, 90\%\}$. 

\end{enumerate}

We create pairwise duplicates between clients such that every client has a unique duplicate element with every other client. All set elements are 32-bit integers. We assume that the \tee can receive data in parallel from the clients. All other interactions that require computation between \tee and clients occur sequentially. In the remainder of this section, by \epmpdone and \epmpdtwo we mean the \epmpd that uses \egpsione and \egpsitwo respectively.

\subsubsection{Effect of Client Count}

\begin{figure*}[t!]
    \centering
    \begin{subfigure}[t]{0.3\textwidth}
        \centering
        \includegraphics[width=1.05\linewidth]{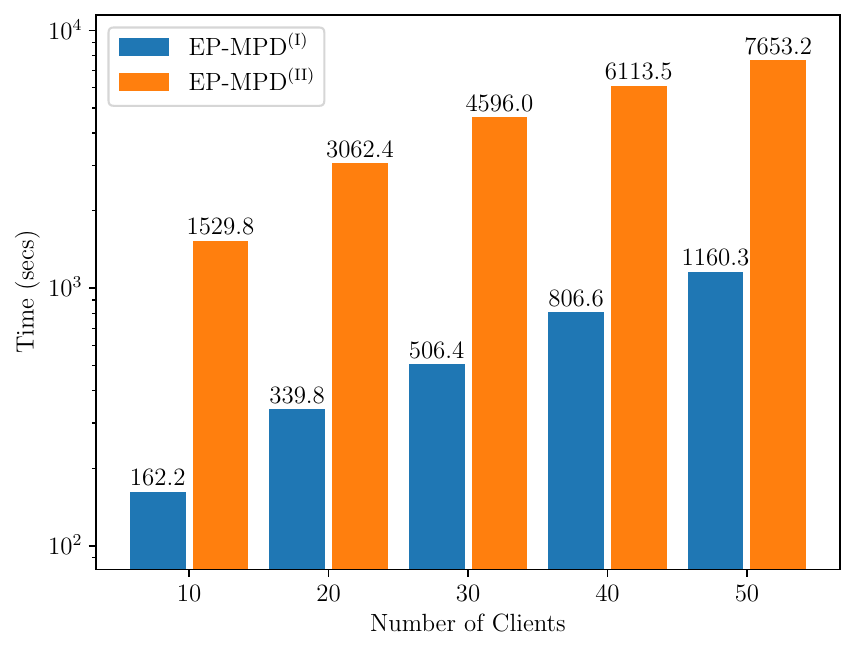}
        \caption{\epmpd running time.}\label{fig_client:tot}
    \end{subfigure}
    ~
%
    \begin{subfigure}[t]{0.3\textwidth}
        \centering
        \includegraphics[width=1.05\linewidth]{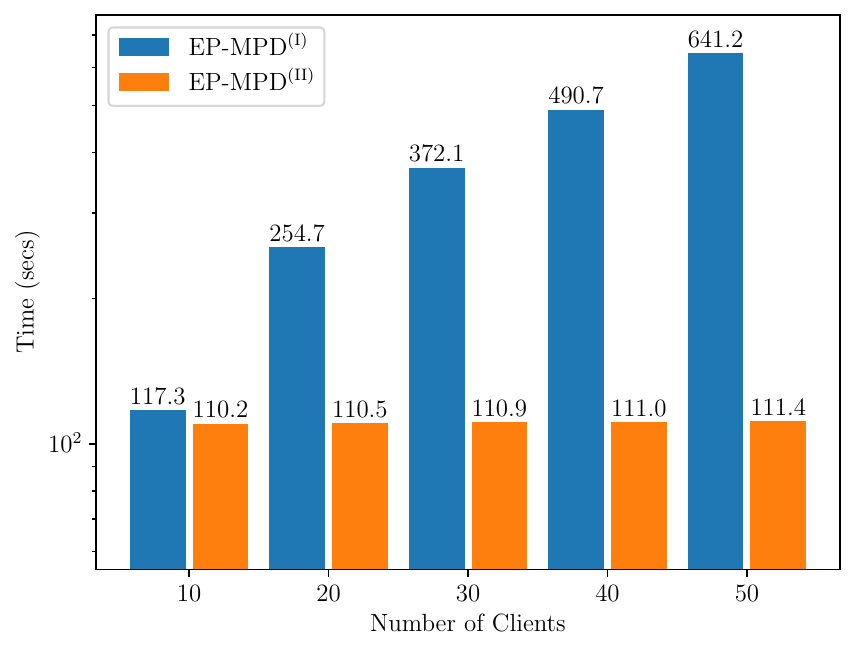}
        \caption{Client running time.}\label{fig_client:client}
    \end{subfigure}
    ~
    \begin{subfigure}[t]{0.3\textwidth}
        \centering
        \includegraphics[width=1.05\linewidth]{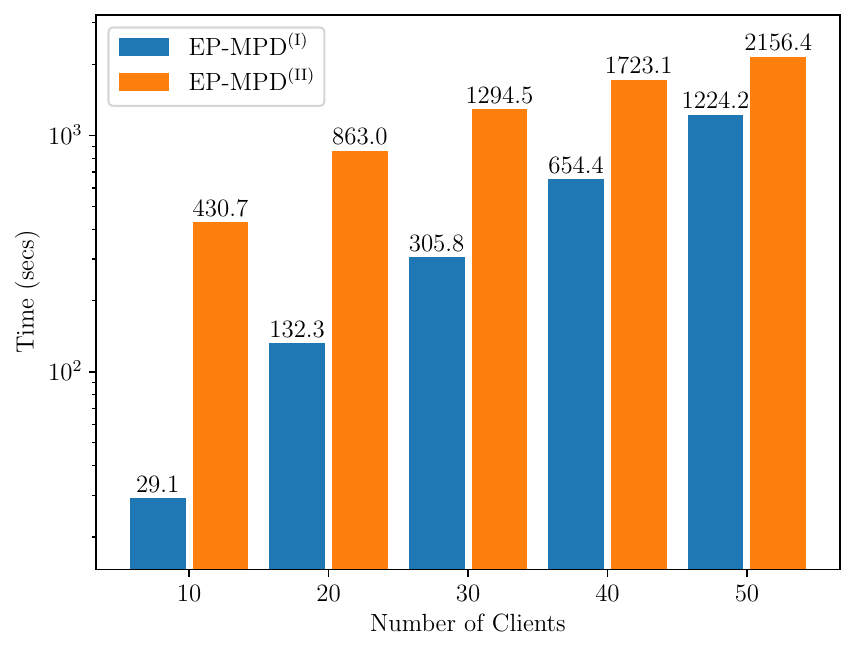}
        \caption{\tee running time.}\label{fig_client:tp}
    \end{subfigure}

    \caption{Effect of client count with $2^{\st 19}$ dataset size and $30\%$ duplication percentage on running time (logarithmic scale).}\label{fig_client}
\end{figure*}

Figure~\ref{fig_client} depicts the effect of the number of clients on (a) \epmpd running time, (b) the average running time of clients, and (c) \tee running time.  Figure~\ref{fig_client:tot} demonstrates that the \epmpd running time increases linearly with the number of clients for both \epmpdone and \epmpdtwo. The increase in the client count rises the number of PRP and OPRF evaluations as the total number of elements across all clients grows. We observe that \epmpdone is about $7\mbox{--}10\times$ faster than \epmpdtwo. 

The average running time of the client however shows a different trend.  The client running time in \epmpdone increases linearly with the client count and is always greater than -- up to $6\times$ -- the running time of \epmpdtwo. The running time of \epmpdtwo is constant as the number of clients increases. In \epmpdone, the client-side running time increases with the number of clients, because each client has to perform more PRP evaluations. However, in \epmpdtwo, the number of OPRF evaluations is equal to the number of elements in each client's set. Additionally, in \epmpdone, the clients perform the  PRP evaluations at every level of the tree. However, in \epmpdtwo the OPRF evaluations are only performed once when the clients are at the leaf level. 
%
The \tee running time in \epmpdtwo is always higher than \epmpdone. Both exhibit linear increases in running time as the number of clients increases. The running time in \epmpdtwo is much higher as the OPRF evaluations are relatively expensive.  

\subsubsection{Effect of Dataset Size}

\begin{figure*}[t!]
    \centering
    \begin{subfigure}[t]{0.3\textwidth}
        \centering
        \includegraphics[width=1.05\linewidth]{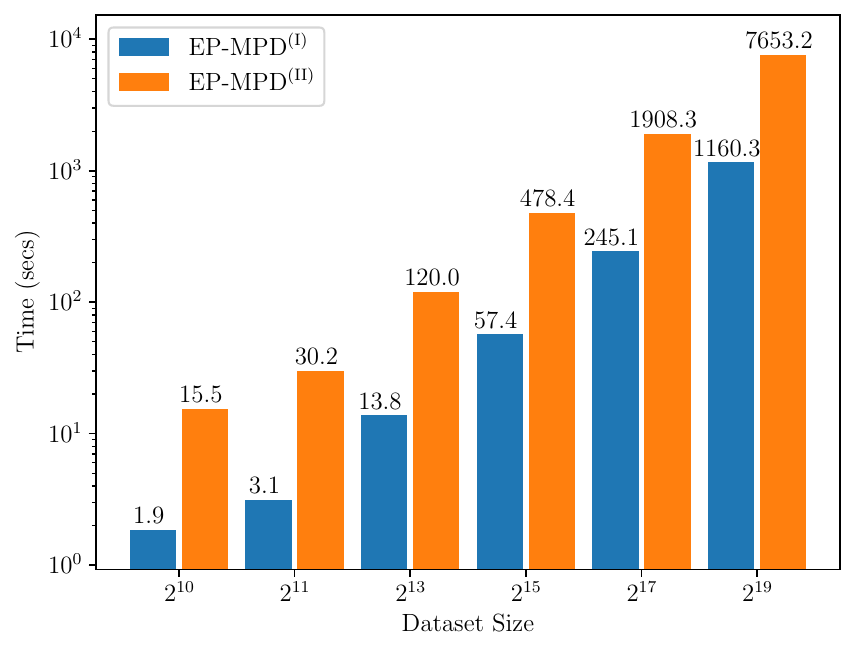}
        \caption{\epmpd running time.}\label{fig_dsize:tot}
    \end{subfigure}
    ~
%
    \begin{subfigure}[t]{0.3\textwidth}
        \centering
        \includegraphics[width=1.05\linewidth]{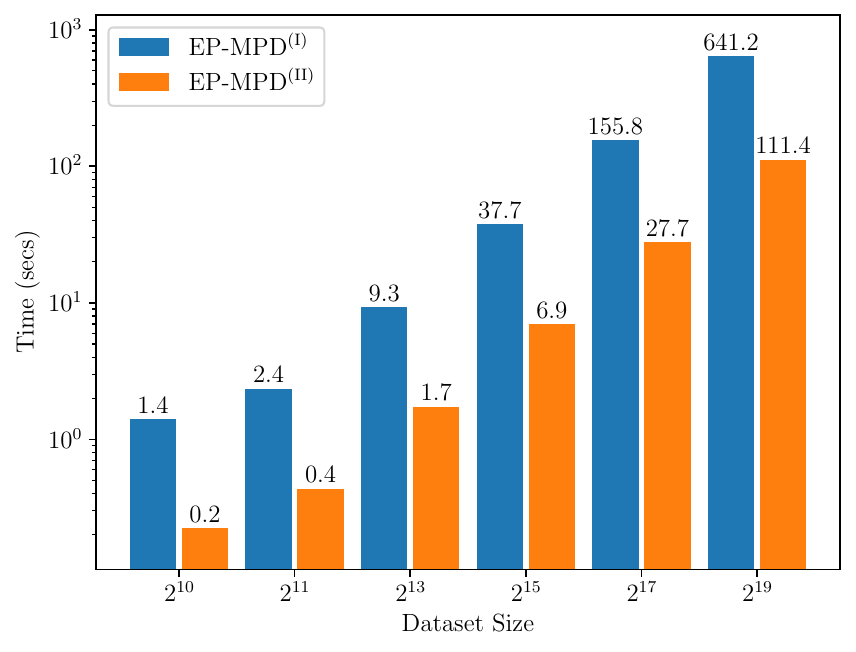}
        \caption{Client running time.}\label{fig_dsize:client}
    \end{subfigure}
    ~
    \begin{subfigure}[t]{0.3\textwidth}
        \centering
        \includegraphics[width=1.05\linewidth]{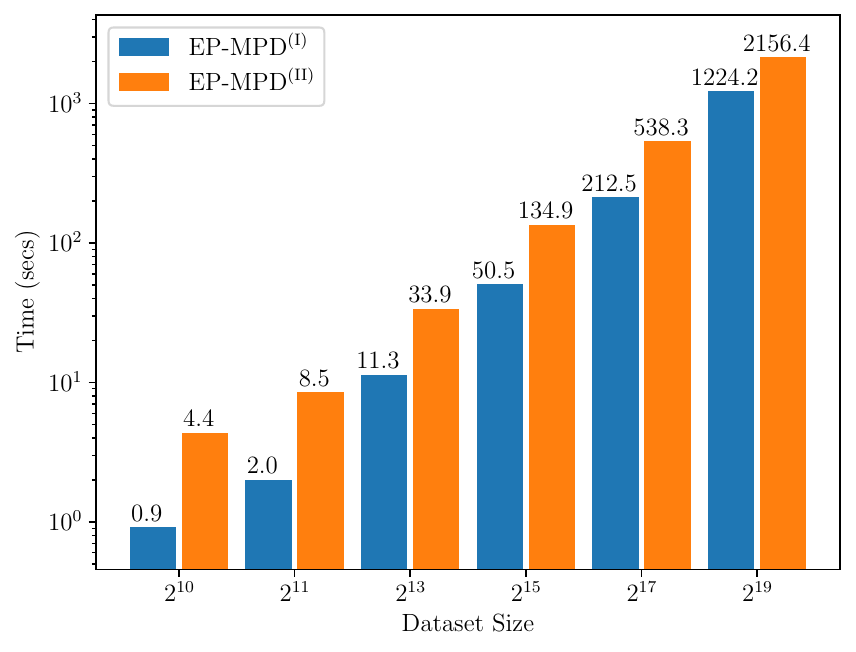}
        \caption{\tee running time.}\label{fig_dsize:tp}
    \end{subfigure}

    \caption{Effect of dataset size with 50 clients and $30\%$ duplication percentage on running time (logarithmic scale).}\label{fig_dsize}
\end{figure*}

Figure~\ref{fig_dsize} shows the effect of dataset size on \epmpd running time, the average running time of clients, and \tee running time. From Figure~\ref{fig_dsize:tot}, it is evident that \epmpd's running time grows linearly with increasing the dataset size. \epmpdone is up to $8\times$ faster than \epmpdtwo. As the dataset size increases, the number of PRP and OPRF evaluations grows, thereby increasing the \epmpd running time. 
Figure~\ref{fig_dsize:client} shows the effect of dataset size on client running time. Unlike increasing the client count, increasing the dataset size increases the client running time of both \epmpdone and \epmpdtwo. We observe this difference because an increase in dataset size increases the number of OPRF evaluations while any change in the client count does not. The client running time of \epmpdtwo is up to $7\times$ faster than \epmpdone. 

Figure~\ref{fig_dsize:tp} depicts the effect of dataset size on the running time of \tee. The growth of dataset size increases the OPRF evaluations in \epmpdtwo. In \epmpdone, \tee has to identify more overlapping ciphertexts as the datasets are larger. The running time of \tee increases linearly with the dataset size for both cases. \tee running time in \epmpdone is up to $5\times$ faster than \epmpdtwo.

\subsubsection{Effect of Duplication Percentage}

\begin{figure*}[t!]
    \centering
    \begin{subfigure}[t]{0.3\textwidth}
        \centering
        \includegraphics[width=1.05\linewidth]{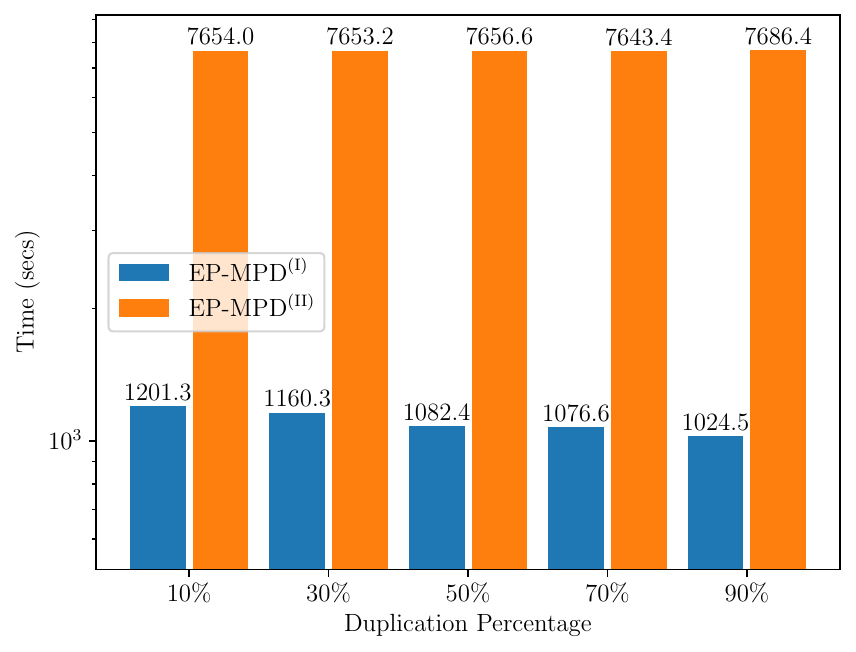}
        \caption{\epmpd running time.}\label{fig_dp:total}
    \end{subfigure}
    ~
%
    \begin{subfigure}[t]{0.3\textwidth}
        \centering
        \includegraphics[width=1.05\linewidth]{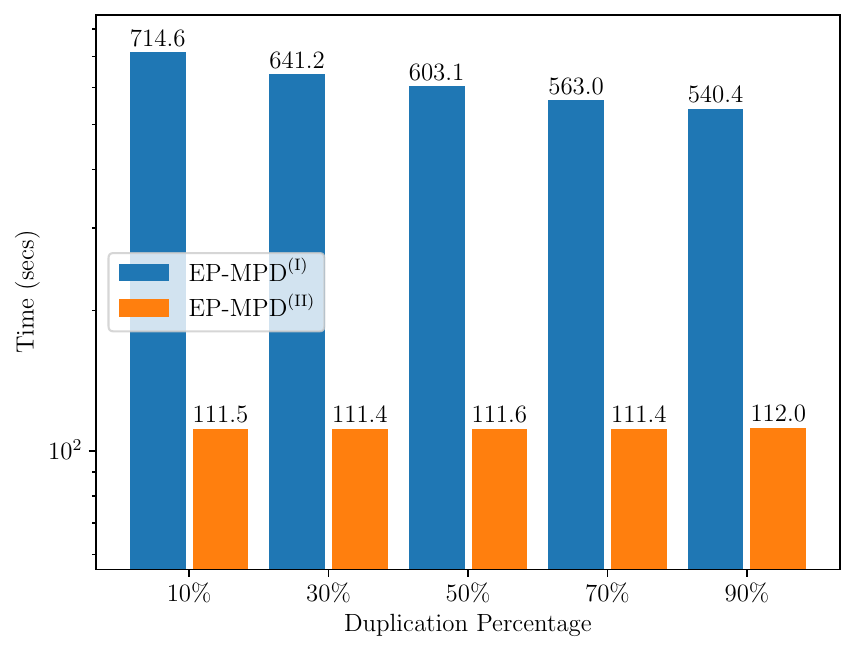}
        \caption{Client running time.}\label{fig_dp:client}
    \end{subfigure}
    ~
    \begin{subfigure}[t]{0.3\textwidth}
        \centering
        \includegraphics[width=1.05\linewidth]{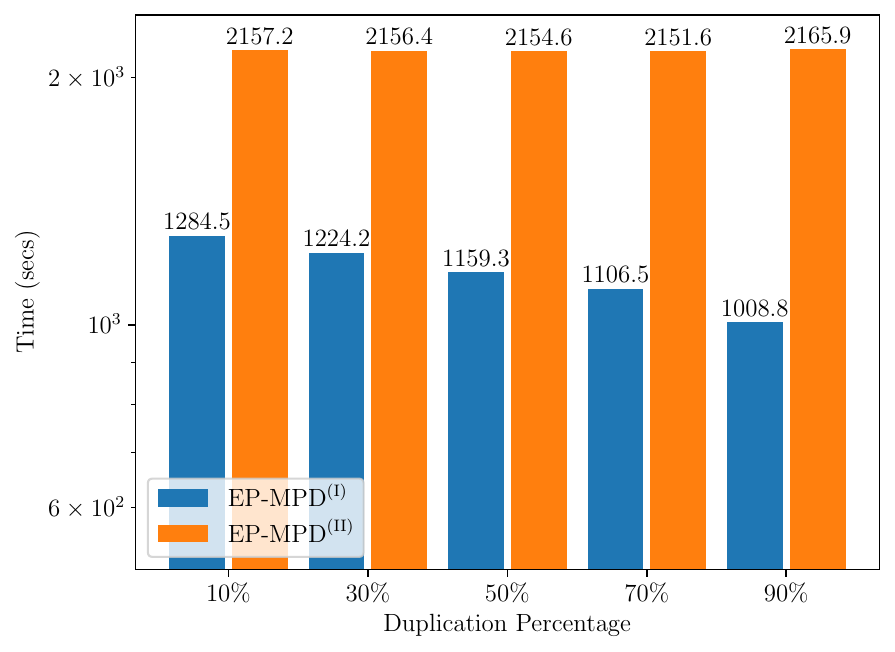}
        \caption{\tee running time.}\label{fig_dp:tp}
    \end{subfigure}

    \caption{Effect of duplication percentage with $2^{\st 19}$ dataset size and $50$ clients on running time (logarithmic scale).}\label{fig_dp}
\end{figure*}

Figure~\ref{fig_dp} shows the effect of duplication percentage on \epmpd running time, the average running time of clients, and \tee running time. Figure~\ref{fig_dp:total} highlights the running time of \epmpd as the duplication percentage changes. We note that in \epmpdtwo, OPRF evaluations incur the most overwhelming running time, dominating all other running time costs. Overall, the running time of \epmpdtwo is constant while that of \epmpdone linearly decreases as the duplication rate grows. In \epmpdtwo, the number of OPRF evaluations is independent of the duplication percentage, because the OPRF is always performed only when the clients are at the leaf level. However, in \epmpdone, the number of PRP invocations decreases as clients move up the tree from the leaf level because a higher duplication percentage removes more elements in each level of the tree. The total running time of \epmpdone is up to $7\times$ faster than \epmpdtwo. 

Furthermore, the client-side running time of \epmpdtwo is constant while that of \epmpdone linearly decreases as the duplication percentage increases, as shown in Figure~\ref{fig_dp:client}. The reason is identical to the observation of the \epmpd running time. The number of PRP evaluations reduces while the number of OPRF evaluations does not. Similar to the prior experiment settings, the client time of \epmpdone is higher than \epmpdtwo and is up to $7\times$ slower. 
Figure~\ref{fig_dp:tp} shows the effect of duplication percentage on the running time of \tee. The \tee running time in \epmpdtwo is constant while in \epmpdone linearly decreases. The \tee running time in \epmpdone is up to $2\times$ faster than \epmpdtwo. 


The overall running time of the symmetric key-based \epmpdone is considerably less than the public key-based \epmpdtwo. Nevertheless, the clients in \epmpdone need to perform more computations than those in \epmpdtwo.

\subsection{\revision{Comparison with Naive Approaches}}
\subsubsection{\revision{Employing Fully Trusted Third Party}}
\label{tee-solution}
\revision{
One naive approach could be exploiting \tee as a full source of trust and \tee collects all clients' sensitive data to build a global model. As we will explain, our approach offers both efficiency and security advantages over this naive method.
%
} 
\revision{
Firstly, the core principle behind FL, including our scheme, is to ensure that clients retain training data within their local devices and share only model updates. Requiring clients to send plaintext data to a \tee conflicts with this principle. While trusted execution environments provide a degree of security, they are not invulnerable—recent research has exposed their susceptibility to side-channel and other types of attacks, making them less trustworthy than previously believed.}

\revision{
Secondly, FL often involves many clients, each holding large datasets. Sending all data to a \tee for model training would demand significant resources. Given the current limitations of trusted execution environments, this approach is inefficient. In contrast, FL and our system enable model training to occur on clients' devices, relieving the computational burden on the \tee.} 
%
%
\revision{
In our schemes, if a semi-honest adversary corrupts \tee, then the adversary will only learn the size of the intersection that each pair of clients from a different group has (as defined in Definition \ref{def::leakage}), and nothing more, as proven in Appendices \ref{sec::proof-of-EG-PSI-1} and \ref{sec::proof-of-EG-PSI-2}. 
Also, \tee never learns secret information about the parties (such as their secret keys). 
}

\subsubsection{\revision{Comparison with Two-Party PSI}}
\label{naive-2psi}

\begin{table}[]
\caption{\revision{Running time (seconds) of \epmpd and \textit{naive} approach using two-party PSI}}
\label{t:brute_force}
\centering
\revision{
\resizebox{\columnwidth}{!}{
\begin{tabular}{|c|cc|c|}
\hline
\multirow{2}{*}{\textbf{Client Count}} & \multicolumn{2}{c|}{\textbf{\epmpd}}                    & \multirow{2}{*}{\textbf{\begin{tabular}[c]{@{}c@{}}Naive Approach \\ (Two-party PSI)\\
\end{tabular}}} \\ \cline{2-3}
                                       & \multicolumn{1}{c|}{\textbf{\epmpdone}} & \textbf{\epmpdtwo} &                                                                                            \\ \hline
10                                     & \multicolumn{1}{c|}{162.2}           & 1529.8          & 616.8                                                                                      \\ \hline
20                                     & \multicolumn{1}{c|}{339.8}           & 3062.4          & 2537.4                                                                                     \\ \hline
30                                     & \multicolumn{1}{c|}{506.4}           & 4596.0          & 5890.8                                                                                     \\ \hline
40                                     & \multicolumn{1}{c|}{806.6}           & 6113.5          & 9884.0                                                                                     \\ \hline
50                                     & \multicolumn{1}{c|}{1160.3}          & 7653.2          & 15974.1                                                                                    \\ \hline
\end{tabular}
}
}
\end{table}

\revision{
As outlined in Section~\ref{naive_approach}, each client can invoke a two-party PSI protocol with every other client to remove all pairwise duplicates. However, such a naive approach would require ${n\choose 2}$ invocations of a two-party PSI for $n$ clients. To highlight the practicality of \epmpd, we implement this naive approach using the highly efficient symmetric-key based PSI proposed in \cite{two_psi}. Table~\ref{t:brute_force} compares our protocols with the naive approach when the data size is $2^{\st 19}$ with $30\%$ duplication. 
%
%
We observe that in all cases \epmpdone outperforms the naive application of two-party PSI. However, the OPRF-based \epmpdtwo is faster for settings with more than 20 clients. The speed-up offered by using \epmpdone over two-party PSI increases as the client count increases. For 10 clients we see an improvement of $\approx 4\times$ and this increases to $\approx 14\times$ for 50 clients.
}

\subsection{\fl with \epmpd}

\subsubsection{Setup}

We create an \fl setting with varying levels of duplication to analyze the effect of \epmpd on the model performance and training efficiency of CLM fine-tuning. We use GPT-2 Medium and GPT-2 Large CLMs from  GPT-2 \cite{radford2019language} family of decoder-only models based on the Transformer \cite{vaswani_attention} architecture.  GPT-2 Medium has 345 million parameters with 24 decoder blocks, 16 attention heads, and 1024 hidden sizes while GPT-2 Large has 774 million parameters with 36 decoder blocks, 20 attention heads, and 1280 hidden sizes. Both models have a sequence length of 1024 tokens. 

We create 10 clients and experiment with the following 7 datasets. The Rotten Tomatoes and IMDB datasets contain movie reviews. The Sonnets and Plays datasets are a collection of works by William Shakespeare. Poetry is a collection of short poems by a variety of poets. Haiku and Short Jokes are a collection of short Japanese-style poems and Jokes from Reddit respectively. The language modeling task for each dataset is to learn how to generate text from the dataset's distribution. For all datasets, we create a Train and Test split in an 80:20 ratio. Model perplexity is evaluated on the Test set after training.  We follow a two-step procedure to create the client datasets with $x \%$ duplication. Initially, we distributed the Train set equally among all 10 clients. Next, we sample with replacement $x\%$ of the original Train set. This subset is then again equally distributed among all clients. We perform 5 rounds of \fl training with at most 10 epochs of local training. The GPT-2 Large models were trained for fewer epochs and with less data because they were overfitting to the Train set for more epochs and data. Additional details about the datasets and \fl training hyperparameters can be found in Appendix~\ref{sec::fl_params}.

\subsubsection{\fl model performance}

We measure the perplexity of the final model on the Test set to assess the impact of various duplication levels on model performance. Table~\ref{t:pp} shows the results of perplexity (PP) and improvement rate (IR). All values in the table are rounded to two decimal places. IR refers to the percentage of improvement in perplexity after deduplication.  For example, the Haiku dataset with GPT-2 Medium and 30\% duplication shows an IR of 4.36\% from 3.78 to 3.61 after deduplication. The deduplicated client datasets represent the baseline perplexity.

The \textcolor{blue}{blue} text in the table highlights the largest improvement in perplexity, which is for the Plays dataset with GPT-2 Medium. We observe an improvement rate of $19.62\%$ from 34.89 to 28.04 for $20\%$ duplication. The Plays dataset in general has a higher perplexity than other datasets as it is a relatively challenging task with longer sequences and fewer training samples. The impact of duplicates in such scenarios is aggravated. We observe a similar trend with the Sonnets dataset with a $13.64\%$ improvement rate for $30\%$ duplication, which is also a difficult task with few samples and relatively long sequences.

The \textcolor{purple}{purple} text highlights the least improvement in perplexity for the Haiku dataset with GPT-2 Medium. The $10\%$ duplication results in a $1.09\%$ improvement rate from $3.65$ to $3.61$. Similarly, the Rotten Tomatoes, Short Jokes, and IMDB datasets exhibit smaller improvement rates across different levels of duplication, while maintaining low perplexity. 
We observe these trends because these datasets are relatively easier to learn with a greater number of training samples. The impact of duplicates in such cases is not as pronounced. 

The \textcolor{red}{red} text highlights two scenarios where deduplication results in insignificant changes ($<1\%)$ to perplexity. The Plays dataset with GPT-2 Medium shows an improvement of $0.3\%$ after deduplication with $10\%$ duplication while the Sonnets dataset with GPT-2 Large worsens by $0.29\%$ after deduplication with $10\%$ duplication. The reason for the ``noisy'' results for the Plays and Sonnets dataset is that both datasets have very few training data samples. A $10\%$ duplication percentage merely adds $3\textup{--}5$ duplicated data points to each client. Both datasets, however, show significant improvements for $20\%$ and $30\%$ duplication. 

\begin{table*}[]
\centering
\caption{Test set perplexity (PP) and improvement rate (IR) of perplexity after deduplication.}
\label{t:pp}
\begin{tabular}{|c|c|cccccc|c|}
\hline
\multirow{3}{*}{\textbf{Model}}                                              & \multirow{3}{*}{\textbf{Dataset}} & \multicolumn{6}{c|}{\textbf{Duplication Percentage}}                                                                                                         & \multirow{2}{*}{\textbf{Deduplicated}} \\ \cline{3-8}
                                                                        &                                   & \multicolumn{2}{c|}{30\%}                                 & \multicolumn{2}{c|}{20\%}                                 & \multicolumn{2}{c|}{10\%}            &                                        \\ \cline{3-9} 
                                                                        &                                   & \multicolumn{1}{c|}{PP}    & \multicolumn{1}{c|}{IR (\%)} & \multicolumn{1}{c|}{PP}    & \multicolumn{1}{c|}{IR (\%)} & \multicolumn{1}{c|}{PP}    & IR (\%) & PP                                     \\ \hline\hline
\multirow{7}{*}{\begin{tabular}[c]{@{}c@{}}GPT-2\\ Medium\end{tabular}} & Haiku                             & \multicolumn{1}{c|}{3.78}  & \multicolumn{1}{c|}{4.36}    & \multicolumn{1}{c|}{3.7}  & \multicolumn{1}{c|}{2.37}     & \multicolumn{1}{c|}{\textcolor{purple}{3.65}}   & \textcolor{purple}{1.09}    & 3.61                                   \\ \cline{2-9} 
                                                                        & Rotten Tomatoes                   & \multicolumn{1}{c|}{2.4}   & \multicolumn{1}{c|}{3.62}    & \multicolumn{1}{c|}{2.36}  & \multicolumn{1}{c|}{2.0}    & \multicolumn{1}{c|}{2.35}  & 1.67     & 2.31                                   \\ \cline{2-9} 
                                                                        & Short Jokes                       & \multicolumn{1}{c|}{3.96}  & \multicolumn{1}{c|}{5.34}    & \multicolumn{1}{c|}{3.89}  & \multicolumn{1}{c|}{3.79}    & \multicolumn{1}{c|}{3.83}  & 2.31    & 3.74                                   \\ \cline{2-9} 
                                                                        & Poetry                            & \multicolumn{1}{c|}{6.24}  & \multicolumn{1}{c|}{14.47}    & \multicolumn{1}{c|}{6.51}  & \multicolumn{1}{c|}{18.07}   & \multicolumn{1}{c|}{5.77}  & 7.61    & 5.33                                   \\ \cline{2-9} 
                                                                        & IMDB                              & \multicolumn{1}{c|}{13.17} & \multicolumn{1}{c|}{7.1}    & \multicolumn{1}{c|}{12.57} & \multicolumn{1}{c|}{2.67}    & \multicolumn{1}{c|}{12.49}  & 2.05     & 12.23                                  \\ \cline{2-9} 
                                                                        & Sonnets                           & \multicolumn{1}{c|}{15.83} & \multicolumn{1}{c|}{13.64}   & \multicolumn{1}{c|}{15.63} & \multicolumn{1}{c|}{12.54}    & \multicolumn{1}{c|}{14.23} & 3.88    & 13.67                                  \\ \cline{2-9} 
                                                                        & Plays                             & \multicolumn{1}{c|}{34.32} & \multicolumn{1}{c|}{18.3}   & \multicolumn{1}{c|}{\textcolor{blue}{34.89}} & \multicolumn{1}{c|}{\textcolor{blue}{19.62}}   & \multicolumn{1}{c|}{\textcolor{red}{28.12}} & \textcolor{red}{--}      & 28.04                                  \\ \hline\hline
\multirow{4}{*}{\begin{tabular}[c]{@{}c@{}}GPT-2\\ Large\end{tabular}}  & Haiku                             & \multicolumn{1}{c|}{3.26}  & \multicolumn{1}{c|}{11.29}    & \multicolumn{1}{c|}{3.25}  & \multicolumn{1}{c|}{10.83}    & \multicolumn{1}{c|}{2.98}  & 2.78    & 2.89                                   \\ \cline{2-9} 
                                                                        & Rotten Tomatoes                   & \multicolumn{1}{c|}{2.65}  & \multicolumn{1}{c|}{16.79}   & \multicolumn{1}{c|}{2.61}   & \multicolumn{1}{c|}{15.29}   & \multicolumn{1}{c|}{2.53}  & 12.81   & 2.21                                    \\ \cline{2-9} 
                                                                        & Short Jokes                       & \multicolumn{1}{c|}{4.11}  & \multicolumn{1}{c|}{7.84}    & \multicolumn{1}{c|}{4.03}  & \multicolumn{1}{c|}{5.86}    & \multicolumn{1}{c|}{3.94}  & 3.64    & 3.79                                   \\ \cline{2-9} 
                                                                        & Sonnets                           & \multicolumn{1}{c|}{8.52}  & \multicolumn{1}{c|}{5.58}    & \multicolumn{1}{c|}{8.4}   & \multicolumn{1}{c|}{4.27}    & \multicolumn{1}{c|}{\textcolor{red}{8.02}}  & \textcolor{red}{--}      & 8.04                                   \\ \hline
\end{tabular}
\end{table*}


\subsubsection{\fl model training efficiency}

We analyze the model training efficiency by computing the total GPU running time of all clients during training. The CPU-bound \epmpd costs are negligible for 10 clients and at most 5,000 elements (the size of the largest dataset is 50,000 elements distributed among all clients). Table~\ref{t:time} highlights the results for GPU running time. All values in the table are rounded to two decimal places. The Time column is the total GPU running time and IR is the percentage improvement compared to the running time within the deduplicated setting.

The \textcolor{blue}{blue} text shows the largest improvement in running time of $27.95\%$ from 33.13 minutes to 23.87 minutes in the GPT-2 Medium experiment with the Sonnets dataset. The \textcolor{purple}{purple} text highlights the smallest improvement in running time of $7.33\%$ from 11.87 to 11.0 minutes in the Sonnets experiment with GPT-2 Large. 
We observe a relatively consistent improvement trend of about $22\%$, $16\%$, and $8\%$ for $30\%$, $20\%$, and $10\%$ duplication levels respectively. The GPT-2 Medium experiments show higher running times as these models were trained longer.  We observe that the running time efficiency consistently improves across all duplication levels and experiments. 

Tables~\ref{t:pp} and  \ref{t:time} show the effectiveness of applying \epmpd to \fl of language models. 
The experiments show improvements of up to $19.62\%$  in perplexity and $27.95\%$  in GPU running time. Deduplication significantly reduces resource usage while enhancing the quality of \fl models. The RTX A6000 GPU used in our experiments consumes 0.3 kWh of energy. 
As \fl scales to larger datasets and more clients, the energy savings with \epmpd become more significant, contributing to more sustainable and cost-effective \fl implementations, benefiting  the environment and the end-users.

\begin{table*}[]
\centering
\caption{Total GPU training time (minutes) of all clients and improvement rate (IR) of time after deduplication.}
\label{t:time}
\begin{tabular}{|c|c|cccccc|c|}
\hline
\multirow{3}{*}{\textbf{Model}}                                         & \multirow{3}{*}{\textbf{Dataset}} & \multicolumn{6}{c|}{\textbf{Duplication Percentage}}                                                                                                               & \multirow{2}{*}{\textbf{Deduplicated}} \\ \cline{3-8}
                                                                        &                                   & \multicolumn{2}{c|}{30\%}                                   & \multicolumn{2}{c|}{20\%}                                   & \multicolumn{2}{c|}{10\%}              &                                        \\ \cline{3-9} 
                                                                        &                                   & \multicolumn{1}{c|}{Time}    & \multicolumn{1}{c|}{IR (\%)} & \multicolumn{1}{c|}{Time}    & \multicolumn{1}{c|}{IR (\%)} & \multicolumn{1}{c|}{Time}    & IR (\%) & Time                                   \\ \hline\hline
\multirow{7}{*}{\begin{tabular}[c]{@{}c@{}}GPT-2\\ Medium\end{tabular}} & Haiku                             & \multicolumn{1}{c|}{111.64}  & \multicolumn{1}{c|}{23.06}   & \multicolumn{1}{c|}{101.67}  & \multicolumn{1}{c|}{15.51}   & \multicolumn{1}{c|}{93.82}   & 8.44    & 85.9                                  \\ \cline{2-9} 
                                                                        & Rotten Tomatoes                   & \multicolumn{1}{c|}{162.79}  & \multicolumn{1}{c|}{21.7}    & \multicolumn{1}{c|}{151.54}  & \multicolumn{1}{c|}{15.89}   & \multicolumn{1}{c|}{138.76}  & 8.14    & 127.46                                 \\ \cline{2-9} 
                                                                        & Short Jokes                       & \multicolumn{1}{c|}{396.62}  & \multicolumn{1}{c|}{27.85}   & \multicolumn{1}{c|}{338.69}  & \multicolumn{1}{c|}{15.51}   & \multicolumn{1}{c|}{313.35}  & 8.68    & 286.15                                 \\ \cline{2-9} 
                                                                        & Poetry                            & \multicolumn{1}{c|}{101.33}  & \multicolumn{1}{c|}{22.55}   & \multicolumn{1}{c|}{94.25}  & \multicolumn{1}{c|}{16.73}    & \multicolumn{1}{c|}{86.87}   & 9.66    & 78.48                                  \\ \cline{2-9} 
                                                                        & IMDB                              & \multicolumn{1}{c|}{2103.93} & \multicolumn{1}{c|}{23.99}   & \multicolumn{1}{c|}{1945.94} & \multicolumn{1}{c|}{17.82}   & \multicolumn{1}{c|}{1772.44} & 9.77    & 1599.24                                \\ \cline{2-9} 
                                                                        & Sonnets                           & \multicolumn{1}{c|}{\textcolor{blue}{33.13}}   & \multicolumn{1}{c|}{\textcolor{blue}{27.95}}   & \multicolumn{1}{c|}{28.53}   & \multicolumn{1}{c|}{16.33}   & \multicolumn{1}{c|}{26.14}   & 8.68    & 23.87                                  \\ \cline{2-9} 
                                                                        & Plays                             & \multicolumn{1}{c|}{31.48}   & \multicolumn{1}{c|}{22.9}    & \multicolumn{1}{c|}{29.38}   & \multicolumn{1}{c|}{17.39}   & \multicolumn{1}{c|}{26.95}   & 9.94    & 24.27                                  \\ \hline\hline
\multirow{4}{*}{\begin{tabular}[c]{@{}c@{}}GPT-2\\ Large\end{tabular}}  & Haiku                             & \multicolumn{1}{c|}{21.0}   & \multicolumn{1}{c|}{22.05}   & \multicolumn{1}{c|}{19.65}   & \multicolumn{1}{c|}{16.69}   & \multicolumn{1}{c|}{18.02}    & 9.16     & 16.37                                  \\ \cline{2-9} 
                                                                        & Rotten Tomatoes                   & \multicolumn{1}{c|}{70.74}   & \multicolumn{1}{c|}{23.08}   & \multicolumn{1}{c|}{65.26}   & \multicolumn{1}{c|}{16.63}   & \multicolumn{1}{c|}{59.91}   & 9.18    & 54.41                                  \\ \cline{2-9} 
                                                                        & Short Jokes                       & \multicolumn{1}{c|}{340.75}  & \multicolumn{1}{c|}{22.93}   & \multicolumn{1}{c|}{313.65}  & \multicolumn{1}{c|}{16.27}   & \multicolumn{1}{c|}{288.86}  & 9.08    & 262.63                                 \\ \cline{2-9} 
                                                                        & Sonnets                           & \multicolumn{1}{c|}{13.91}   & \multicolumn{1}{c|}{20.92}   & \multicolumn{1}{c|}{12.89}   & \multicolumn{1}{c|}{14.66}   & \multicolumn{1}{c|}{\textcolor{purple}{11.87}}   & \textcolor{purple}{7.33}    & 11.0                                   \\ \hline
\end{tabular}
\end{table*}


\section{Asymptotic Costs Analysis}\label{sec:Asymptotic-Cost-Analysis} 

In this section,  we analyze the asymptotic costs of our schemes. Table \ref{PSI-concrete-communication} summarizes the result.



\begin{table*}[!htb]
\begin{center}
\caption{ \small  Complexities of our solutions; namely, \egpsione, \egpsitwo, and \epmpd. In the table, we have considered the maximum communication  complexity of a client in \epmpd (without considering the cases where the updates reduce the clients' set size). In the table, $m$ is the number of clients in each group, $|\set|$ is the maximum cardinality of sets, and $|R|$ is the maximum cardinality of sets intersection that a client has with clients of the other group. 
}\label{PSI-concrete-communication}

\renewcommand{\arraystretch}{1.4}
\scalebox{0.9}{
\begin{tabular}{|c|c|c|c|c|c|c|c|c|c|c|c|c|c|c|} 
   \hline
 \multirow{2}{*} { \textbf{Protocols}}&

 \multicolumn{2}{c|}{  \textbf{Computation Cost}}&\multicolumn{2}{c|}{  \textbf{Communication Cost}}\\

 \cline{2-5}

  &  Client&   $\tee$&  Client&   \tee  \\
  

      \hline
 {\rotatebox[origin=c]{0}{   \egpsione}}&\cellcolor{gray!20}   $O(m \cdot |\set_{\st j, i}|)$&\cellcolor{gray!20}   $O(m \cdot |\set|)$&\cellcolor{gray!20}  $O(m \cdot |S_{\st j, i}|)$ &\cellcolor{gray!20}  $O(m \cdot|R|)$\\

     \hline
   
 {\rotatebox[origin=c]{0}{   \egpsitwo}}&\cellcolor{gray!50}  $O(m \cdot |\set|)$&\cellcolor{gray!50}  $O(m\cdot |S|)$&\cellcolor{gray!50}  $O(m\cdot |\set_{\st j, i}|)$ &\cellcolor{gray!50}  $O(m\cdot |\set|)$\\
  
 \hline

 {\rotatebox[origin=c]{0}{   \epmpd}}&\cellcolor{gray!20}  $O(m\cdot\log_{\st 2}(m) \cdot |S|)$&\cellcolor{gray!20}  $O(m \cdot \log_{\st 2}(m) \cdot |S|)$&\cellcolor{gray!20}  $O(m\cdot \log_{\st 2}(m) \cdot |\set_{\st j, i}|)$ &\cellcolor{gray!20}  $O(m\cdot \log_{\st 2}(m) \cdot |\set|)$\\
  
 \hline
 
\end{tabular}
}
\end{center}
\end{table*}

\subsection{Costs of \egpsione and \egpsitwo}\label{sec:PSI-Asymptotic-Cost-Analysis}

\subsubsection{Computation Cost} Initially, we focus on the computation cost of parties in \egpsione. In step \ref{EG-PSI-1-enc-elem}, each client $\cl_{\st j, i}$ encrypts each element of its set using the symmetric key encryption \PRP, under each key it agrees with the clients in the other group. Specifically, it invokes \PRP $m \cdot |\set_{\st j, i}|$ times, where $m$ is the total number of clients in the other group. In step \ref{EG-PSI-1-extract-res}, each  client $\cl_{\st j, i}$ invokes \PRP $m \cdot |R_{\st j, i}|$ times, where $R_{\st j, i}$ is the (encrypted) intersection set. Thus, the computation complexity of each $\cl_{\st j, i}$ is $O(m \cdot |\set_{\st j, i}|)$. 
The computation complexity of \tee in step \ref{EG-PSI-1-server-side} is $O(m \cdot |\set|)$, where $|\set|$ represents the maximum cardinality of sets. The only computation that \tee performs is searching to find duplicated values. 

Next, we analyze the computation cost of parties in \egpsitwo. In step \ref{EG-PSI-2-enc-set}, each  $\cl_{\st j, i}$ invokes $|\set_{\st j, i}|$ times OPRF to encrypt its set elements. In steps \ref{EG-PSI-2-find-intersection} and \ref{EG-PSI-2-extract}, each $\cl_{\st j, i}$ performs a search on $m+1$ encrypted sets, which imposes negligible costs compared to the executions of OPRF. Thus, the complexity of each $\cl_{\st j, i}$ is $O(m \cdot |\set|)$, where $|\set|$ is the maximum cardinality of sets. \tee invokes an instance of OPRF in step \ref{EG-PSI-2-enc-set} for every set of elements of each client. Thus, its computation complexity is $O(m\cdot |S|)$, where $|S|$ is the maximum cardinality of sets.

\subsubsection{Communication Cost}  

We first focus on the communication cost of parties in \egpsione. Each  $\cl_{\st 0, i}$ in step \ref{EG-PSI-1-setup-key-agr} transmits $m$ keys to the clients of the other group. Thus, its communication complexity in this step is $O(m)$. However, the clients in $\gr_{\st 1}$ do not transmit any message in this step.  
In step \ref{EG-PSI-1-enc-elem}, each  $\cl_{\st j, i}$ of each group, sends $m \cdot |\set_{\st j, i}|$ messages to \tee, where the size of each message is only about $256$ bits. Thus, each client's communication complexity is $O(m \cdot |S_{\st j, i}|)$.  \tee sends to each client $R_{\st j,i}$ which contains the intersection the client has with the clients in the other group. Thus, the overall communication complexity of \tee is $O(m \cdot|R|)$, where $|R|$ is the maximum cardinality of sets intersection that a client may have with clients of the other group.  

Now, we evaluate the communication complexity of parties in \egpsitwo. In step \ref{EG-PSI-2-enc-set}, each $\cl_{\st j, i}$ performs OPRF evaluations on $\set_{\st j, i}$. This procedure imposes $O(|\set_{\st j, i}|)$ communication cost to the client. In step \ref{EG-PSI-2-find-intersection}, the communication complexity of each $\cl_{\st j, i}$ is $O(m\cdot|\set_{\st j, i}|)$ because it sends its encrypted set to every client in the other group. In step \ref{EG-PSI-2-enc-set}, \tee performs OPRF evaluations on the elements of each $\cl_{\st j, i}$ using OPRF. This procedure imposes $O(m\cdot |\set|)$ communication cost to \tee.

\subsection{Costs of \epmpd}\label{sec:DE-Asymptotic-Cost-Analysis} 

\subsubsection{Computation Cost}
In step \ref{DE-find-intersection}, at each level $d$, each $\cl_{\st j, i}$ invokes an instance of \egpsi. When \egpsione or \egpsitwo is used, then the computation complexity of $\cl_{\st j, i}$ in this step is $O(m\cdot\log_{\st 2}(m) \cdot |\set|)$, where $|\set|$ is the maximum cardinality of the clients' sets. 
In step \ref{DE-remove-duplicates}, $\cl_{\st j, i}$ updates its set. This involves simply removing items from its local set and the associated cost is negligible compared to other costs. Thus, the overall computation complexity of $\cl_{\st j, i}$ is $O(m\cdot\log_{\st 2}(m) \cdot |S|)$. The computation cost is dominated by the cost of executing \PRP and OPRF if \egpsione and \egpsitwo are used respectively. The computation complexity of \tee in either case (when \egpsione or \egpsitwo is used) is $O(m \cdot \log_{\st 2}(m) \cdot |S|)$. 


\subsubsection{Communication Cost} In step \ref{DE-find-intersection}, at each level $d$, each client $\cl_{\st 0, i}$ transmits its updated set to an instance of \egpsione or \egpsitwo. Therefore, its overall communication complexity is $O\Big(m\cdot \big(|\set_{\st 0, i}| + \sum\limits_{\st u = 2}^{\st d}(|\set_{\st 0, i}|-\sum\limits_{\st u' = 1}^{\st u - 1}|\set_{\st \cap}^{\st (u')}|)\big)\Big)$ where $d = \log_{\st 2}(m)$ and $|\set_{\st \cap}^{\st (u')}|$ is the size of the intersection that $\cl_{\st 0, i}$ has with the clients of the other group at level $u'$. On the other hand, at each level $d$, the  communication complexity of each client $\cl_{\st 1, i}$ is $O(m\cdot \log_{\st 2}(m) \cdot |\set_{\st 1, i}|)$. Note that the majority of the clients will eventually become members of group 0. Thus, their set size eventually reduces when they (i) proceed to a higher level in the tree and (ii) have intersections with other clients' sets.

The communication complexity of \tee, when \egpsione is used, is $O(m\cdot \log_{\st 2}(m) \cdot|R|)$, where $|R|$ is the maximum cardinality of sets intersection that a client may have with clients of the other group. The communication complexity of \tee, when \egpsitwo is used, is 
%
%
$O(m\cdot \log_{\st 2}(m) \cdot |\set|)$. 

\section{Related Works}
A detailed deduplication method has been proposed in \cite{lee-etal-2022-deduplicating} that used existing techniques of finding duplicates such as suffix array \cite{MM93}  and MinHash \cite{B97}. However, there were no privacy requirements in this method as the data owner is responsible for local deduplication. As we consider the problem of deduplication in a distributive setting, thus our problem statement is fundamentally different from \cite{lee-etal-2022-deduplicating}.

In \cite{KBR13}, the problem of privacy-preserving deduplication by a centralized service was discussed -- there is a server that stores multiple users' files after removing duplicates. Their method involves a second server which helps in setting up keys for each user using an OPRF. Our work is different from \cite{KBR13} in many ways. First, we apply OPRF for encrypting the dataset elements, which allows efficient deduplication. Second, our deduplication requirement is more granular -- we want entries of users' datasets to be deduplicated. If the two datasets have the same elements irrespective of their order, then at the end of our deduplication, one of the users will receive an empty dataset and the other will receive their dataset unchanged. On the other hand, since \cite{KBR13} considers a file as a whole (not every set element), so two files $M_{\st 1}$ and $M_{\st 2}$ have the same elements but in a different order will not be removed. Third, our public key-based deduplication protocol is federated as opposed to \cite{KBR13} which is centralized (all data is stored in a centralized server responsible for the deduplication). Fourth, \cite{KBR13} is a two-server setting -- one server helps in deduplication and the other stores the deduplicated files. In contrast, in our protocol apart from the users, we only have one third-party.

Another work along the lines of \cite{KBR13} is  \cite{LNOPV20} which proposed Multi-Key Revealing Encryption (MKRE) that lets a server compress the encrypted files of multiple users by removing ciphertexts related to similar plaintexts. Their construction is based on message lock encryption, therefore, it is subject to dictionary attacks \cite{cryptoeprint:2020/799}. Our scheme is not vulnerable to such attacks and, as a result, provides a stronger security guarantee.

The work of \cite{BK23} also has pointed out the need for data processing for ML and applied PSI-type tools. However, their problem statement is to find a mismatch in the datasets belonging to two participants. In contrast, we aim to find duplicates (matching data) in sets belonging to two or more parties. Additionally, their application domain is narrow as they only considered labeled data. Their end goal is to find entries in two different sets where their features match but their corresponding labels do not. On the other hand, our protocol supports both labeled and unlabeled data.

We discuss other related work on PSI protocols in Appendix~\ref{sec::related-work}.



\section{Conclusions}
\label{sec::conclusion}
It was established in \cite{lee-etal-2022-deduplicating} that machine learning benefits significantly from deduplication. However, it was not evident how federated learning (FL) could benefit from deduplication without compromising data privacy. Our successful and efficient solution to this problem has paved the way for further applications. 
Our pioneering protocols, \egpsione and \egpsitwo, can be employed in any distributed system to identify and efficiently remove duplicates. Although this paper considers elements as duplicates only if they are exact copies, there are scenarios where elements that are not exact matches but are sufficiently similar should also be treated as duplicates \cite{lee-etal-2022-deduplicating}. Our current primitives can easily be extended to address this case. By applying fuzzy hash functions, which map closely related elements to the same digest, the deduplication of near-match elements can be reduced to the problem of deduplication of exact-match elements, allowing the use of \egpsione and \egpsitwo in distributed setups. While our system's application focuses on text-based learning, it can also be extended to other areas, such as image processing and speech recognition, thereby broadening its utility and impact.

\section*{Acknowledgments}

   Aydin Abadi was supported in part by EPSRC under grants EP/Y028813/1 and EP/W003325/1. Sumanta Sarkar acknowledges EPSRC grants EP/T014784/1 and EP/X036669/1.

\bibliographystyle{IEEEtran}
\bibliography{IEEEabrv,mlbib}

\begin{thebibliography}{10}
\providecommand{\url}[1]{#1}
\csname url@samestyle\endcsname
\providecommand{\newblock}{\relax}
\providecommand{\bibinfo}[2]{#2}
\providecommand{\BIBentrySTDinterwordspacing}{\spaceskip=0pt\relax}
\providecommand{\BIBentryALTinterwordstretchfactor}{4}
\providecommand{\BIBentryALTinterwordspacing}{\spaceskip=\fontdimen2\font plus
\BIBentryALTinterwordstretchfactor\fontdimen3\font minus
  \fontdimen4\font\relax}
\providecommand{\BIBforeignlanguage}[2]{{%
\expandafter\ifx\csname l@#1\endcsname\relax
\typeout{** WARNING: IEEEtran.bst: No hyphenation pattern has been}%
\typeout{** loaded for the language `#1'. Using the pattern for}%
\typeout{** the default language instead.}%
\else
\language=\csname l@#1\endcsname
\fi
#2}}
\providecommand{\BIBdecl}{\relax}
\BIBdecl

\bibitem{ICbook19}
I.~F. Ilyas and X.~Chu, \emph{Data Cleaning}.\hskip 1em plus 0.5em minus
  0.4em\relax New York, NY, USA: Association for Computing Machinery, 2019.

\bibitem{lee-etal-2022-deduplicating}
K.~Lee, D.~Ippolito, A.~Nystrom, C.~Zhang, D.~Eck, C.~Callison-Burch, and
  N.~Carlini, ``Deduplicating training data makes language models better,'' in
  \emph{Proceedings of the 60th Annual Meeting of the Association for
  Computational Linguistics}, 2022.

\bibitem{Raffel-C4}
C.~Raffel, N.~Shazeer, A.~Roberts, K.~Lee, S.~Narang, M.~Matena, Y.~Zhou,
  W.~Li, and P.~J. Liu, ``Exploring the limits of transfer learning with a
  unified text-to-text transformer,'' \emph{J. Mach. Learn. Res.}, 2020.

\bibitem{carlini2023quantifyingmemorizationneurallanguage}
\BIBentryALTinterwordspacing
N.~Carlini, D.~Ippolito, M.~Jagielski, K.~Lee, F.~Tramer, and C.~Zhang,
  ``Quantifying memorization across neural language models,'' 2023. [Online].
  Available: \url{https://arxiv.org/abs/2202.07646}
\BIBentrySTDinterwordspacing

\bibitem{suri2023subjectmembershipinferenceattacks}
\BIBentryALTinterwordspacing
A.~Suri, P.~Kanani, V.~J. Marathe, and D.~W. Peterson, ``Subject membership
  inference attacks in federated learning,'' 2023. [Online]. Available:
  \url{https://arxiv.org/abs/2206.03317}
\BIBentrySTDinterwordspacing

\bibitem{mattern2023membership}
J.~Mattern, F.~Mireshghallah, Z.~Jin, B.~Sch{\"{o}}lkopf, M.~Sachan, and
  T.~Berg{-}Kirkpatrick, ``Membership inference attacks against language models
  via neighbourhood comparison,'' in \emph{{ACL}}, 2023.

\bibitem{nasr2023scalableextractiontrainingdata}
M.~Nasr, N.~Carlini, J.~Hayase, M.~Jagielski, A.~F. Cooper, D.~Ippolito, C.~A.
  Choquette-Choo, E.~Wallace, F.~Tramèr, and K.~Lee, ``Scalable extraction of
  training data from (production) language models,'' 2023.

\bibitem{carlini21extracting}
N.~Carlini, F.~Tramer, E.~Wallace, M.~Jagielski, A.~Herbert-Voss, K.~Lee,
  A.~Roberts, T.~Brown, D.~Song, U.~Erlingsson, A.~Oprea, and C.~Raffel,
  ``Extracting training data from large language models,'' in \emph{USENIX
  Security}, 2021.

\bibitem{karamolegkou2023copyrightviolationslargelanguage}
\BIBentryALTinterwordspacing
A.~Karamolegkou, J.~Li, L.~Zhou, and A.~Søgaard, ``Copyright violations and
  large language models,'' 2023. [Online]. Available:
  \url{https://arxiv.org/abs/2310.13771}
\BIBentrySTDinterwordspacing

\bibitem{kandpal2022deduplicatingtrainingdatamitigates}
\BIBentryALTinterwordspacing
N.~Kandpal, E.~Wallace, and C.~Raffel, ``Deduplicating training data mitigates
  privacy risks in language models,'' 2022. [Online]. Available:
  \url{https://arxiv.org/abs/2202.06539}
\BIBentrySTDinterwordspacing

\bibitem{rashid2024fltrojan}
\BIBentryALTinterwordspacing
M.~R.~U. Rashid, V.~A. Dasu, K.~Gu, N.~Sultana, and S.~Mehnaz, ``Fltrojan:
  Privacy leakage attacks against federated language models through selective
  weight tampering,'' 2024. [Online]. Available:
  \url{https://arxiv.org/abs/2310.16152}
\BIBentrySTDinterwordspacing

\bibitem{BGMS21}
E.~M. Bender, T.~Gebru, A.~McMillan{-}Major, and S.~Shmitchell, ``On the
  dangers of stochastic parrots: Can language models be too big?'' in
  \emph{FAccT}, 2021.

\bibitem{SGM19}
E.~Strubell, A.~Ganesh, and A.~McCallum, ``Energy and policy considerations for
  deep learning in {NLP},'' in \emph{{ACL}}, 2019.

\bibitem{Patterson-etal-carbon19}
D.~A. Patterson, J.~Gonzalez, Q.~V. Le, C.~Liang, L.~Munguia, D.~Rothchild,
  D.~R. So, M.~Texier, and J.~Dean, ``Carbon emissions and large neural network
  training,'' \emph{CoRR}, 2021.

\bibitem{fedml_intro}
H.~B. McMahan, E.~Moore, D.~Ramage, and B.~A. y~Arcas, ``Federated learning of
  deep networks using model averaging,'' \emph{CoRR}, 2016.

\bibitem{ACKAE22}
R.~S. Antunes, C.~A. da~Costa, A.~K{\"u}derle, I.~A. Yari, and B.~Eskofier,
  ``Federated learning for healthcare: Systematic review and architecture
  proposal,'' \emph{TIST}, 2022.

\bibitem{sensor}
J.~C. Jiang, B.~Kantarci, S.~Oktug, and T.~Soyata, ``Federated learning in
  smart city sensing: Challenges and opportunities,'' \emph{Sensors}, 2020.

\bibitem{XYTWL2021}
Q.~Xia, W.~Ye, Z.~Tao, J.~Wu, and Q.~Li, ``A survey of federated learning for
  edge computing: Research problems and solutions,'' \emph{High-Confidence
  Computing}, 2021.

\bibitem{google_keyboard}
A.~Hard, K.~Rao, R.~Mathews, F.~Beaufays, S.~Augenstein, H.~Eichner, C.~Kiddon,
  and D.~Ramage, ``Federated learning for mobile keyboard prediction,''
  \emph{CoRR}, 2018.

\bibitem{DBLP:conf/eurocrypt/FreedmanNP04}
M.~J. Freedman, K.~Nissim, and B.~Pinkas, ``Efficient private matching and set
  intersection,'' in \emph{{EUROCRYPT}}, 2004.

\bibitem{Adensidora-abadi}
A.~Abadi, ``Earn while you reveal: Private set intersection that rewards
  participants,'' \emph{CoRR}, 2023.

\bibitem{AbadiDMT22}
A.~Abadi, C.~Dong, S.~J. Murdoch, and S.~Terzis, ``Multi-party updatable
  delegated private set intersection,'' in \emph{{FC}}, 2022.

\bibitem{fowl2023decepticonscorruptedtransformersbreach}
\BIBentryALTinterwordspacing
L.~Fowl, J.~Geiping, S.~Reich, Y.~Wen, W.~Czaja, M.~Goldblum, and T.~Goldstein,
  ``Decepticons: Corrupted transformers breach privacy in federated learning
  for language models,'' 2023. [Online]. Available:
  \url{https://arxiv.org/abs/2201.12675}
\BIBentrySTDinterwordspacing

\bibitem{balunovic2022lampextractingtextgradients}
\BIBentryALTinterwordspacing
M.~Balunović, D.~I. Dimitrov, N.~Jovanović, and M.~Vechev, ``Lamp: Extracting
  text from gradients with language model priors,'' 2022. [Online]. Available:
  \url{https://arxiv.org/abs/2202.08827}
\BIBentrySTDinterwordspacing

\bibitem{cheng2021secureboost}
K.~Cheng, T.~Fan, Y.~Jin, Y.~Liu, T.~Chen, D.~Papadopoulos, and Q.~Yang,
  ``Secureboost: A lossless federated learning framework,'' \emph{IEEE
  Intelligent Systems}, 2021.

\bibitem{provfl}
V.~A. Dasu, S.~Sarkar, and K.~Mandal, ``Prov-fl: Privacy-preserving round
  optimal verifiable federated learning,'' ser. AISec'22, 2022.

\bibitem{Xu_2019}
R.~Xu, N.~Baracaldo, Y.~Zhou, A.~Anwar, and H.~Ludwig, ``Hybridalpha,''
  \emph{AISec}, 2019.

\bibitem{bonawitz_new}
J.~H. Bell, K.~A. Bonawitz, A.~Gasc\'{o}n, T.~Lepoint, and M.~Raykova, ``Secure
  single-server aggregation with (poly)logarithmic overhead,'' ser. CCS, 2020.

\bibitem{BIKMMPRS17}
K.~Bonawitz, V.~Ivanov, B.~Kreuter, A.~Marcedone, H.~B. McMahan, S.~Patel,
  D.~Ramage, A.~Segal, and K.~Seth, ``Practical secure aggregation for
  privacy-preserving machine learning,'' in \emph{{CCS}}, 2017.

\bibitem{dp_sgd}
M.~Abadi, A.~Chu, I.~Goodfellow, H.~B. McMahan, I.~Mironov, K.~Talwar, and
  L.~Zhang, ``Deep learning with differential privacy,'' ser. CCS, 2016.

\bibitem{DBLP:books/cu/Goldreich2004}
O.~Goldreich, \emph{The Foundations of Cryptography - Volume 2, Basic
  Applications}.\hskip 1em plus 0.5em minus 0.4em\relax Cambridge University
  Press, 2004.

\bibitem{CCXZLL20}
G.~Chen, S.~Chen, Y.~Xiao, Y.~Zhang, Z.~Lin, and T.~Lai, ``Sgxpectre: Stealing
  intel secrets from {SGX} enclaves via speculative execution,'' \emph{{IEEE}
  Secur. Priv.}, 2020.

\bibitem{DMM23}
F.~D{\"{o}}rre, J.~Mechler, and J.~M{\"{u}}ller{-}Quade, ``Practically
  efficient private set intersection from trusted hardware with
  side-channels,'' in \emph{{ASIACRYPT}}, J.~Guo and R.~Steinfeld, Eds., 2023.

\bibitem{PassST17}
R.~Pass, E.~Shi, and F.~Tram{\`{e}}r, ``Formal abstractions for attested
  execution secure processors,'' in \emph{{EUROCRYPT}}, 2017.

\bibitem{ZhangZ16}
F.~Zhang and H.~Zhang, ``Sok: {A} study of using hardware-assisted isolated
  execution environments for security,'' in \emph{HASP}, 2016.

\bibitem{TramerZLHJS17}
F.~Tram{\`{e}}r, F.~Zhang, H.~Lin, J.~Hubaux, A.~Juels, and E.~Shi,
  ``Sealed-glass proofs: Using transparent enclaves to prove and sell
  knowledge,'' in \emph{EuroS{\&}P}, 2017.

\bibitem{burns_oprf}
J.~Burns, D.~Moore, K.~Ray, R.~Speers, and B.~Vohaska, ``Ec-oprf: Oblivious
  pseudorandom functions using elliptic curves,'' Cryptology ePrint Archive,
  2017.

\bibitem{two_psi}
S.~Kamara, P.~Mohassel, M.~Raykova, and S.~Sadeghian, ``Scaling private set
  intersection to billion-element sets,'' in \emph{Financial Cryptography and
  Data Security}, N.~Christin and R.~Safavi-Naini, Eds.\hskip 1em plus 0.5em
  minus 0.4em\relax Berlin, Heidelberg: Springer Berlin Heidelberg, 2014, pp.
  195--215.

\bibitem{radford2019language}
A.~Radford, J.~Wu, R.~Child, D.~Luan, D.~Amodei, and I.~Sutskever, ``Language
  models are unsupervised multitask learners,'' 2019.

\bibitem{vaswani_attention}
\BIBentryALTinterwordspacing
A.~Vaswani, N.~Shazeer, N.~Parmar, J.~Uszkoreit, L.~Jones, A.~N. Gomez,
  L.~Kaiser, and I.~Polosukhin, ``Attention is all you need,'' \emph{CoRR},
  vol. abs/1706.03762, 2017. [Online]. Available:
  \url{http://arxiv.org/abs/1706.03762}
\BIBentrySTDinterwordspacing

\bibitem{MM93}
U.~Manber and E.~W. Myers, ``Suffix arrays: a new method for on-line string
  searches,'' \emph{SIAM J. Comput.}, 1993.

\bibitem{B97}
A.~Z. Broder, ``On the resemblance and containment of documents,'' in
  \emph{{SEQUENCES}}, 1997.

\bibitem{KBR13}
S.~Keelveedhi, M.~Bellare, and T.~Ristenpart, ``Dupless: Server-aided
  encryption for deduplicated storage,'' in \emph{{USENIX} Security}, 2013.

\bibitem{LNOPV20}
D.~E. Lucani, L.~Nielsen, C.~Orlandi, E.~Pagnin, and R.~Vestergaard, ``Secure
  generalized deduplication via multi-key revealing encryption,'' in
  \emph{{SCN}}, 2020.

\bibitem{cryptoeprint:2020/799}
\BIBentryALTinterwordspacing
------, ``Secure generalized deduplication via multi-key revealing
  encryption,'' Cryptology ePrint Archive, Paper 2020/799, 2020. [Online].
  Available: \url{https://eprint.iacr.org/2020/799}
\BIBentrySTDinterwordspacing

\bibitem{BK23}
E.~Blass and F.~Kerschbaum, ``Private collaborative data cleaning via non-equi
  psi,'' in \emph{IEEE SP}, 2023.

\bibitem{RaghuramanR22}
S.~Raghuraman and P.~Rindal, ``Blazing fast {PSI} from improved {OKVS} and
  subfield {VOLE},'' in \emph{{CCS}}, 2022.

\bibitem{DorreMM23}
F.~D{\"{o}}rre, J.~Mechler, and J.~M{\"{u}}ller{-}Quade, ``Practically
  efficient private set intersection from trusted hardware with
  side-channels,'' in \emph{{ASIACRYPT}}.\hskip 1em plus 0.5em minus
  0.4em\relax Springer, 2023.

\bibitem{DBLP:conf/scn/InbarOP18}
R.~Inbar, E.~Omri, and B.~Pinkas, ``Efficient scalable multiparty private
  set-intersection via garbled bloom filters,'' in \emph{{SCN}}, 2018.

\bibitem{DBLP:conf/ccs/KolesnikovMPRT17}
V.~Kolesnikov, N.~Matania, B.~Pinkas, M.~Rosulek, and N.~Trieu, ``Practical
  multi-party private set intersection from symmetric-key techniques,'' in
  \emph{{CCS}}, 2017.

\bibitem{Ben-EfraimNOP22}
A.~Ben{-}Efraim, O.~Nissenbaum, E.~Omri, and A.~Paskin{-}Cherniavsky,
  ``Psimple: Practical multiparty maliciously-secure private set
  intersection,'' in \emph{{ASIA} {CCS}}, 2022.

\bibitem{GhoshN19}
S.~Ghosh and T.~Nilges, ``An algebraic approach to maliciously secure private
  set intersection,'' in \emph{{EUROCRYPT}}, 2019.

\bibitem{ZhangLLJL19}
E.~Zhang, F.~Liu, Q.~Lai, G.~Jin, and Y.~Li, ``Efficient multi-party private
  set intersection against malicious adversaries,'' in \emph{{CCSW}}, 2019.

\bibitem{NevoTY21}
O.~Nevo, N.~Trieu, and A.~Yanai, ``Simple, fast malicious multiparty private
  set intersection,'' in \emph{{CCS}}, 2021.

\bibitem{kamarascaling}
S.~Kamara, P.~Mohassel, M.~Raykova, and S.~Sadeghian, ``Scaling private set
  intersection to billion-element sets,'' in \emph{{FC}}, 2014.

\bibitem{imdb}
A.~L. Maas, R.~E. Daly, P.~T. Pham, D.~Huang, A.~Y. Ng, and C.~Potts,
  ``Learning word vectors for sentiment analysis,'' in \emph{Proceedings of the
  49th Annual Meeting of the Association for Computational Linguistics: Human
  Language Technologies}, 2011.

\bibitem{loshchilov2019decoupledweightdecayregularization}
\BIBentryALTinterwordspacing
I.~Loshchilov and F.~Hutter, ``Decoupled weight decay regularization,'' 2019.
  [Online]. Available: \url{https://arxiv.org/abs/1711.05101}
\BIBentrySTDinterwordspacing

\bibitem{smc_dp}
J.~B{\"o}hler and F.~Kerschbaum, ``Secure multi-party computation of
  differentially private median,'' in \emph{USENIX Security}, 2020.

\end{thebibliography}

\appendices

\section{Security Proof of \egpsione}\label{sec::proof-of-EG-PSI-1}


Next, we prove the security of \egpsione, Theorem~\ref{thm:egpsi-1}.

\begin{proof} We consider cases where in each case a party is corrupt.

\subsection{Corrupt Client}

In this case, for simplicity, we focus on the view of a client $\cl_{\st 1, i}$ in $\gr_{\st 1}$. 
Because the clients of both groups behave the same way, with the main difference being that a client in $\gr_{\st 0}$ generates a key and sends it to each client of $\gr_{\st 1}$. 
Thus, the proof asserting the protocol is secure regarding a corrupt client in $\gr_{\st 1}$ will imply the security of the same protocol concerning a client in $\gr_{\st 0}$.  
%
In the real execution of the protocol, the view of a client, $\cl_{\st 1, i}$, is defined as follows: 
$$\{\mathsf{View}_{\st \cl_{\st 1, i}}^{\st \adv, \egpsione}(S, \emptyset) \}_{\st S} = \{S_{\st 1, i}, r_{\st  1, i}, \vec{k}, R_{\st 1, i}, \vec{v}_{\st 1, i}\}, $$

where $S$ is the set containing all clients' sets, $r_{\st 1,i}$  is the outcome of internal random coins of $\cl_{\st 1, i}$, $\vec{k}=[k_{\st i,1},\ldots, k_{\st i, m}]$ is a vector of $m$ keys received from the clients of $\gr_{\st 0}$, $R_{\st 1, i}$ is the encrypted intersection $\cl_{\st 1, i}$ received from \tee, and $\vec{v}_{\st j,i}$ equals $[\vec{v}_{\st j,i, 1},\ldots, \vec{v}_{\st j,i, m}]$, each $l$-th vector in $\vec{v}_{\st j,i}$ contains the elements that are common between  $\cl_{\st 1, i}$'s set and the set that belongs to the $l$-th client in the other group.  

%
We proceed to construct a simulator, $\simm_{\st \cl_{\st 1, i}}$, in the ideal model. The simulator, receiving the client's input $S_{\st 1, i}$ and output $\vec{v}_{\st 1, i}$, operates as follows. 

\begin{enumerate}
     
    \item initiates an empty view. It appends to it $S_{\st 1, i}$ and uniformly random coins $r'_{\st 1, i}$. It constructs an empty vector $\vec{k}'$ and then chooses $m$ keys for \PRP. It appends the keys to $\vec{k}'$ and adds $\vec{k}'$ to the view.  
    
     \item constructs an empty set $R'$. It encrypts the elements of every $l$-th vector $\vec{v}_{\st 1,i, l}\in \vec{v}_{\st 1,i}$ using $l$-th key $k_{\st l}$ in $\vec{k}'$ and \PRP. It inserts the ciphertexts into $R'$. 
    
    \item appends $R'$ and $\vec{v}_{\st 1,i}$ to the view. It outputs the view. 
\end{enumerate}

Now, we are prepared to discuss that the two views are computationally indistinguishable. The input set $S_{\st 1, i}$ has identical distribution in both models, as their elements are identical. 
Also, since we are in the semi-honest model, in the real execution of the protocol, the client picks its random coins $r_{\st 1, i}$ according to the protocol description. In the ideal model, coins $r'_{\st 1, i}$  have been selected uniformly at random. Therefore $r_{\st 1, i}$ and $r'_{\st 1, i}$ have identical distributions. Vectors $\vec{k}$ and $\vec{k}'$ have identical distributions too because their elements have been chosen uniformly at random. 
In the real model, $R_{\st 1,i}$ contains the encrypted elements of $\vec{v}_{\st 1,i}$, where the keys in $\vec{k}$ are used for the encryption by applying the \PRP. Similarly, in the ideal model, each element of $R'$ is the encryption of an element in $\vec{v}_{\st 1, i}$ where a key in $\vec{k}'$ is used for the encryption with the same \PRP. Thus, $R$ and $R'$ have identical distributions. Also, elements of $\vec{v}_{\st 1, i}$ are identical in both models; therefore, they have identical distributions in the real and ideal models.  
We conclude that the views in the real and ideal models are indistinguishable. 
It is worth noting that the above proof can be easily extended to the case where $n-1$ clients collude with each other. In this case, their \textit{joint} views in the real and ideal models are computationally indistinguishable.

\subsection{Corrupt \tee}

In the real model, the view of \tee is as follows: 
$$\{\mathsf{View}_{\st \tp}^{\st \adv, \Gamma}(S, \emptyset) \}_{\st S}=\{(\set'_{\st 0, 1},\ldots, \set'_{\st 0, m}), (\set'_{\st 1, 1},\ldots,\set'_{\st 1, m})\}$$

where each $\set'_{\st j, i}$ contains the encrypted set elements of client $\cl_{\st j,i}$. Next, we construct a simulator $\simm^{\st \leak}_{\tee}$ that receives the output $\vec{s}$ of leakage function \leak (as defined in Definition \ref{def::leakage}). 

\begin{enumerate}
    \item initiates an empty view. 
    \item constructs $m$ empty sets for each $\gr_{\st 0}$ and $\gr_{\st 1}$. Specifically, it constructs empty sets  $(\underbrace{\set''_{\st 0, 1},\ldots, \set''_{\st 0, m}}_{\st \gr_{\st 0}}), (\underbrace{\set''_{\st 1, 1},\ldots,\set''_{\st 1, m}}_{\st \gr_{\st 1}})$. 

    \item fills each set $\set''_{\st j, i}$ as follows, given that $\vec{s}= [\vec{s}_{\st 0, 1},\ldots, \vec{s}_{\st 0, m}, \vec{s}_{\st 1, 1},$ $\ldots, \vec{s}_{\st 1, m}]$, where  $\vec{s}_{\st j, i}=\Big[\big|[\set_{\st j, i}\cap \set_{\st 1-j, 1}]\big|,\ldots,\big|[\set_{\st j, i}\cap \set_{\st 1-j, m}]\big|\Big]$, as defined in Definition \ref{def::pair-wise-intersection-cardinally}. 
    \begin{enumerate}
        \item for every $y$-th element of each $\vec{s}_{\st 0, i}$ (where $1\leq i,y\leq m$) picks  $\vec{s}_{\st 0, i}[y]$ random values (from the range of \PRP). It appends these values to both sets  $\set''_{\st 0,i}$ and $\set''_{\st 1,y}$. 
        \item pads every set $\set''_{\st j, i}$ with some random values  (from the range of \PRP) such that the size of each $\set''_{\st j, i}$ after padding equals the size of  $\set'_{\st j, i}$.

    \end{enumerate}

  \item appends sets $({\set''_{\st 0, 1},\ldots, \set''_{\st 0, m}}_{\st }), ({\set''_{\st 1, 1}, \ldots,\set''_{\st 1, m}})$ to the view and outputs the view. 
    
\end{enumerate}

We show that the views in the real and ideal models are  indistinguishable. In the real model, each elements of a set $S'_{\st j, i}$ is an output of \PRP. However, in the ideal model, each element of a set $S''_{\st j, i}$ has been picked uniformly at random (from the range of \PRP). By the security of \PRP (i.e., an output of \PRP is computationally indistinguishable from a random value),  sets' elements in the real and ideal model are computationally indistinguishable. 
The size of the intersection between each  set $S'_{\st 0, i}$ and each set $S'_{\st 1, l}$ (in the real model) equals the size of the intersection between each set $S''_{\st 0, i}$  and each set $S''_{\st 1, l}$ (in the ideal model). Also, the size of each $S'_{\st j, i}$ in the real model equals the size of the corresponding set $S''_{\st j, i}$ in the ideal model. Hence, sets  $\set'_{\st 0, 1},\ldots, \set'_{\st 0, m}, \set'_{\st 1, 1},\ldots,\set'_{\st 1, m}$ in the real model are indistinguishable from sets $\set''_{\st 0, 1}, \ldots, \set''_{\st 0, m}, \set''_{\st 1, 1}, \ldots,\set''_{\st 1, m}$ in the ideal model. 
Thus, the views in the real and ideal models are computationally indistinguishable. 
\end{proof}


\section{Security Proof of \egpsitwo}\label{sec::proof-of-EG-PSI-2}

\begin{proof}[Proof (sketch)] 
It is not hard to simulate parties' views. \tee's view in the real model can be easily simulated given that the main information it learns within the protocol execution includes its view during the invocation of the OPRF instances (but not the outputs of OPRF).  If OPRF is secure, then the views of \tee within the real and ideal model will be computationally indistinguishable in this protocol.

Each client's view (in the real execution of the protocol) mainly includes the outputs of OPRF received from its counterparts and the exchanged messages during OPRF's execution with \tee. 
Due to the security of OPRF, an output of OPRF is computationally indistinguishable from a value picked uniformly at random from the output range of OPRF. Also, the messages each client receives during the execution of OPRF can be simulated, due to the security of OPRF. 
\end{proof}

\section{Security Proof of \epmpd}\label{sec::proof-of-DE}

In this section, we prove the security of \epmpd, i.e., Theorem \ref{theorem::securit-of-DE}. For the sake of concreteness and because we have provided a detailed proof for \egpsione, we will prove Theorem \ref{theorem::securit-of-DE} when \epmpd relies on \egpsione. The proof for \egpsione-based \epmpd will easily follow.

%

\begin{proof}
In the real execution of the protocol, the view of a client, $\cl_{\st  i}$, is defined as follows: 
$\{\mathsf{View}_{\st \cl_{\st i}}^{\st \adv, \epmpd}(S) \}_{\st S} = \{S_{\st i},  \vec{\mathsf{View}}_{\st \cl_{\st i}}, \vec{v}_{\st i}\} $, 
where $S = \{\set_{\st 1}, \ldots, \set_{\st  m}\}$, $\set_{\st i}$ represents a set belonging to client $\cl_{\st i}$, $\vec{\mathsf{View}}_{\st \cl_{\st i}}=[\mathsf{View}_{\st \cl_{\st i}, 1}^{\st \adv, \egpsione}, \ldots, \mathsf{View}_{\st \cl_{\st i}, z}^{\st \adv, \egpsione}]$ contains $z = log_{\st 2}(m)$ views of $\cl_{\st  i}$ such that each view $\mathsf{View}_{\st \cl_{\st i}, d}^{\st \adv, \egpsione}$ corresponds to $\cl_{\st  i}$'s view when it invokes \egpsione at level $d$, where $1\leq d \leq z$. Also,  $\vec{v}_{\st i}$ contains $z$ vectors, where each vector $\vec{v}_{\st i, d}$ in $\vec{v}_{\st i}$ is the output of \egpsione that client $\cl_{\st i}$ receives at level $d$. 
We construct a simulator, $\simm_{\st \cl_{\st i}}$, in the ideal model. The simulator, receives the output $Q_{\st i}$ of leakage function \leakW (as defined in Definition \ref{def::leakage-W}), the client's input $S_{\st i}$ and output $\vec{v_{\st i}}$. $\simm_{\st \cl_{\st i}}$ operates as follows. 

\begin{enumerate}

 

 \item\label{sim-ed-gen-set-for-other-group}initiates an empty view. For each level $d$ ($1 \leq d\leq z$) constructs the following set for each client $\cl_{\st l}$ in the other group, using (i) the result $\vec{v}_{\st i, d}$ that $\cl_{\st i}$ receives at level $d$ and (ii) the leakage function's output $Q_{\st i}$.

 \begin{enumerate}
     \item sets an empty set $S_{\st l}$. It finds each triple $(x, a, b)$ in $Q_{\st i}$, where $a = i$ and $b = l$. It adds the first 
     element $x$ of each triple that meets the above conditions to $S_{\st l}$. This ensures that $S_{\st l}$ contains all the elements in common with $S_{\st i}$. 

     \item pads every $\set_{\st l}$ with  random values  (from the set universe) such that the size of each $\set_{\st l}$, in the ideal model, after padding equals the size of $\set_{\st l}$ in the real model. 

 \end{enumerate}

     \item\label{sim-ed-gen-set-for-adv-group} generates $\frac{2^{\st d}}{2}-1$ empty sets for the group to which $S_{\st i}$ belongs. It appends to these empty sets values picked uniformly at random from the set universe. It ensures the size of each of these sets in the ideal model equals to that of in the real model.

     \item for every level $d$: 
     
     \begin{enumerate}
         \item invokes an instance of \egpsione using the following sets: (i) $S_{\st i}$
         and sets generated in step \ref{sim-ed-gen-set-for-adv-group}, that are placed in one group and (ii) sets generated in step \ref{sim-ed-gen-set-for-other-group}, placed in another group. As a result, it receives output $\vec{v}_{\st i, d}$.
         

         \item extracts    $\mathsf{View}_{\st \cl_{\st i}, d}^{\st \adv, \egpsione}$ after \egpsione's execution. 

         \item appends every $\mathsf{View}_{\st \cl_{\st i}, d}^{\st \adv, \egpsione}$ and $\vec{v}_{\st i, d}$ to its view. 

         \item outputs the view. 
     \end{enumerate}

\end{enumerate}

Now, we show that the two views  are  indistinguishable. The input $S_{\st i}$ in both models are identical, therefore they will have identical distribution. 
Due to the security of \egpsione, for every level $d$, $\mathsf{View}_{\st \cl_{\st i}, d}^{\st \adv, \egpsione}$ in the real and ideal models are computationally indistinguishable, as demonstrated in the proof of Theorem \ref{thm:egpsi-1}. Also, the intersection $\vec{v}_{\st i, d}$ that \adv learns at every level $d$ in the real and ideal model are identical, thus they are indistinguishable. 
Hence, the views of \adv in the real and ideal models are computationally indistinguishable. 
\end{proof}



\section{Related Work in PSI Research}\label{sec::related-work}

Since their introduction in \cite{DBLP:conf/eurocrypt/FreedmanNP04}, numerous PSIs have been designed. In general, PSIs fall into two broad categories \textit{traditional} and \textit{delegated}.  
In traditional PSIs, clients (i.e., data owners) compute the result interactively using their data stored locally.
In this research line, researchers in  \cite{RaghuramanR22} introduced two two-party PSIs, one secure against semi-honest and the other against malicious (or active) adversaries. These protocols are the fastest two-party PSIs to date. They use  Oblivious Key-Value Stores (OKVS) data structure and Vector Oblivious Linear Evaluation (VOLE). These solutions' computation cost is $O(c)$, where $c$ is a set's cardinality.  They also impose $O(c\log c^{\st 2}+\kappa)$ and $O(c\cdot \kappa)$ communication costs in the semi-honest and malicious models respectively, 
where 
$\kappa$ is a security parameter.  

Dorre et al. \cite{DorreMM23} proposed an efficient two-party PSI that requires each party to locally posses a secure hardware, a requirement that may not always be possible to meet. The scheme further relies on a hash function and deterministic symmetric-key encryption. The scheme's communication complexity is $O(poly(\kappa)+|S|\lambda)$ and its computation complexity is $O((2\cdot |S|)\cdot poly(\kappa\cdot l))$, where $l$ is a set element's bit size and $\lambda$ is a security parameter. As the authors stated, the protocol cannot directly support multiple more than two parties.

Furthermore, researchers developed PSIs that allows multiple (i.e., more than two) parties to compute the intersection. The multi-party PSIs in  \cite{DBLP:conf/scn/InbarOP18,DBLP:conf/ccs/KolesnikovMPRT17} are secure against semi-honest adversaries while those proposed in \cite{Ben-EfraimNOP22,GhoshN19,ZhangLLJL19,DBLP:conf/ccs/KolesnikovMPRT17,NevoTY21} were developed to remain secure against active ones. 
%
%
The protocols in   \cite{DBLP:conf/ccs/KolesnikovMPRT17} and  \cite{NevoTY21} are the most  efficient multi-party PSIs  designed to be secure against passive and active adversaries respectively. The computation and communication complexities of the PSI in  \cite{DBLP:conf/ccs/KolesnikovMPRT17} are  $O(c\cdot m^{\st 2}+c\cdot m )$ and $O(c\cdot m^{\st 2})$. 
%
%
The PSI in \cite{NevoTY21} has a parameter $t$ that determines how many parties can collude with each other and must be set before the protocol's execution, where $t\in [2, m)$.  
%
%
Its computation and communication complexities are $O(c\cdot \kappa(m+t^{\st 2}-t(m+1)))$ and $O(c\cdot m\cdot \kappa)$ respectively.



Delegated PSIs use cloud computing for computation and/or storage. These PSIs can be further categorized into those that allow \textit{one-off} and \textit{repeated} delegation of PSI computation.  The most efficient one that supports one-off delegation is \cite{kamarascaling}, designed for the two-party setting, with an overhead of $O(c)$.    
Researchers also proposed a multi-party PSI that supports repeated delegation and efficient \emph{updates} \cite{AbadiDMT22}. It imposes $O(h\cdot d^{\st 2}\cdot m)$ and $O(h\cdot d\cdot m)$  computation and communication costs respectively, during the PSI computation. Furthermore, to incentivize clients to participate in a PSI protocol and share their sensitive inputs, researchers proposed a PSI that rewards participants \cite{Adensidora-abadi}.

Thus, despite the development of numerous two-party and multiple-party PSIs, none of them to date can match the features offered by our PSIs.



\section{Datasets and \fl Hyperparameters}\label{sec::fl_params}

We use the following datasets in our study:

\begin{enumerate}
    \item \textit{Haiku}\footnote{\url{https://www.kaggle.com/datasets/bfbarry/haiku-dataset}}: This dataset contains 15,281 Haikus scraped from ``reddit.com/r/haiku''. Haiku is a Japanese style of short poetry.
    \item \textit{Rotten Tomatoes}\footnote{\url{https://huggingface.co/datasets/cornell-movie-review-data/rotten_tomatoes}}:  It contains 10,662 movie reviews from the Rotten Tomatoes movie critic website. Originally intended for sentiment analysis, the dataset has 5,331 positive and negative reviews each. In our experiments, we ignore the sentiment of the reviews and train the CLM only on the review.
    \item \textit{Short Jokes}\footnote{\url{https://www.kaggle.com/datasets/abhinavmoudgil95/short-jokes}}:  It contains 231,657 jokes scraped from Reddit. We randomly sample 50,000 jokes for our experiments. 
    \item \textit{Poetry}\footnote{\url{https://huggingface.co/datasets/merve/poetry}}: It contains 573 poems from the Renaissance and Modern eras by various poets.
    
    \item \textit{IMDB} \cite{imdb}: This dataset contains 50,000 movie reviews from the IMDB movie critic website. Similar to the Rotten Tomatoes dataset, the IMDB dataset was originally intended for sentiment analysis. We ignore the sentiment of each review and only train the CLM on the review.
    \item \textit{Sonnets}\footnote{\url{https://huggingface.co/datasets/Lambent/shakespeare_sonnets_diffused}}: This dataset contains 460 sonnets (including variations) from William Shakespeare. Sonnets are 14-line poems with variable rhyme schemes
    \item \textit{Plays}\footnote{\url{https://huggingface.co/datasets/Trelis/tiny-shakespeare}}: It contains 521 plays from William Shakespeare. 
\end{enumerate}

We train the GPT-2 Medium and GPT-2 Large models for $5\text{--}10$ and $1\text{--}2$ local epochs respectively. Both models are trained for 5 rounds. The Haiku GPT-2 Large was trained for 1 epoch while the other GPT-2 Large models were trained for 2 epochs. The Poetry and Sonnets GPT-2 Medium models were trained for 10 epochs while the others were trained for 5 epochs. We observe that training GPT-2 Large for more epochs and data results in overfitting and poor Test set perplexity across all duplication levels.

For all experiments, we use a learning of $5 \times 10^{\st -5}$ with the AdamW \cite{loshchilov2019decoupledweightdecayregularization} optimizer. We use a linear warm-up schedule with 10\% of the training steps as warm-up steps. We use $\ell_{\st 2}$-norm clipping for the gradients and set the threshold to 1.0. We set the sequence length for the datasets as follows:

\begin{enumerate}[nolistsep]
    \item Haiku: 100
    \item Rotten Tomatoes: 200
    \item Short Jokes: 100
    \item Poetry: 1000
    \item IMDB: 500
    \item Sonnets: 400
    \item Plays: 1000
\end{enumerate}

For the \textit{Short Jokes} dataset, we randomly sample 50,000 elements to ensure the running time is feasible as we train with 10 clients. We set the batch size to values in the range $[4, 32]$, depending on the sequence length and model size to ensure that training can be performed on one RTX A6000 GPU. We use a Train/Test split of 80/20 for all datasets. In the case of datasets like IMDB that have a separate Test set, we combine the original Train and Test sets into a single dataset, shuffle it, and then create 80:20 splits. We fix the random seed to 123 for the Numpy, PyTorch, and Python random number generators to ensure the reproducibility of our results.

\section{\revision{Proof of Correctness of the federated learning with deduplication}}\label{sec::Proof-of-Correctness}

\revision{
In this section, we prove the correctness of the protocol presented in Figure \ref{fig:fedml-dedup-protocol}, i.e., federated learning with deduplication. Informally, a protocol is correct, if it terminates and produces the expected result (i.e., a global model trained on deduplicated databases) in the presence of honest parties. 
\begin{theorem}\label{theorem::correctness-of-dedup-FL}
The ``Federated learning with deduplication'' scheme (presented in Figure \ref{fig:fedml-dedup-protocol}) is correct if the underlying federated learning scheme, the local duplication scheme, and  \epmpd (presented in Figure \ref{fig::construction-PPMD}) are correct. 
\end{theorem}
Before we prove Theorem \ref{theorem::correctness-of-dedup-FL}, we first prove the correctness of (i) \egpsi (i.e., \egpsione or \egpsitwo) which is a component we have developed that forms the foundation for \epmpd, and (ii) \epmpd, which is another component we have developed and upon which the federated learning with deduplication scheme relies. 
\begin{claim}\label{claim::proof-of-EGPSI-correctness}
\egpsione  and \egpsitwo are correct, if \PRP and OPRF are correct respectively. 
\end{claim}
}

\begin{claimproof}
    Initially, we prove \egpsione's correctness. Each client in $\gr_{\st 0}$ with each client in $\gr_{\st 1}$ agrees on an identical key and encrypts their set elements using \PRP and that key. Due to \PRP's correctness (i.e., its deterministic feature), if these clients have elements in common, their encryption will be identical. Thus,  \tee in Phase \ref{EG-PSI-1-server-side} will find the common encrypted elements of the two clients and inform them about these elements. 
    Furthermore, due to the correctness of \PRP and because each pair of clients (each belonging to a different group) that share an identical key initially store triple $(e'_{\st i,l}, k_{\st i,l}, l)$, given a common encrypted element, say $e'_{\st i,l}$, they can extract each related key, e.g., $k_{\st i,l}$ and decrypt the associated ciphertext, (i.e., $\PRP^{\st -1}(k_{\st i, l}, e'_{\st i,l})\rightarrow e$) to identify in its local database the plaintext elements common in both clients' databases.  Thus, they can find the intersection of the two databases.

Now, we prove the correctness of \egpsitwo. Due to the correctness of OPRF (i.e., the deterministic nature of it) and because \tee uses an identical secret key as an input of OPRF for all clients, if any client has an intersection with other clients, the output of OPRF for that element will be identical as well. This allows each client to find the encryption of identical elements in Phase \ref{EG-PSI-2-find-intersection}. Since each client knows which plaintext element of its set corresponds to which ciphertext, it can identify the plaintext elements that are in common with each client of the other group. Thus, it can find the intersection of its set and its counterpart's set.  
\end{claimproof}

\begin{claim}\label{claim::correctness-of-epmpd}
    \epmpd is correct if \egpsi is correct. 
\end{claim}

\begin{claimproof}
We initially focus on the leaf node level. We know that at this level, (1) each group contains  a single client and (2) \egpsi is applied to a pair of groups. Therefore, in Phase \ref{DE-find-intersection}, due to the correctness of  \egpsi (i.e., Claim \ref{claim::proof-of-EGPSI-correctness}), each client can always find the elements it has in common with its counterpart, i.e., the duplicated elements in these two clients' sets. Hence, the client in $\gr_{\st 0}$ can always delete these duplicated elements in Phase \ref{DE-remove-duplicates}. At any level above the leaf node level, (1) the clients in the same group will not have any elements in common, because their intersection has already been identified and removed at the lower levels and (2) due to the correctness of \egpsi (i.e., Claim \ref{claim::proof-of-EGPSI-correctness}) the intersection of the sets of each client in one group with each client of the other group is identified by these two clients. Thus, each client in $\gr_{\st 0}$ can delete the duplicated element in Phase \ref{DE-remove-duplicates}, at any level above the leaf node level as well.  Moreover, since only a client in $\gr_{\st 0}$ deletes the identified duplicated elements, a copy of that element still remains in the set of $\gr_{\st 1}$, meaning that union of $\gr_{\st 0}$'s set and $\gr_{\st 1}$'s set is the same after this stage. 
\end{claimproof}
\begin{proof}
We prove Theorem \ref{theorem::correctness-of-dedup-FL}. Due to the correctness of a local deduplication mechanism (that can simply find replicated elements in a single dataset and remove the repeated elements) each client will not have any repeated elements by the end of Phase \ref{Local-Deduplication}. Also, due to the correctness of \epmpd (i.e., Claim \ref{claim::correctness-of-epmpd}),  
by the end of Phase \ref{Global-Deduplication}, there will be no duplicated elements across all clients' data. \epmpd also ensures the union of all clients' datasets is the same after the completion of this deduplication. FL runs on the union of all clients' datasets, hence, due to the FL's correctness (discussed in Section \ref{sec::FL}) the protocol will complete and all parties will always learn the final global model, trained on the deduplicated datasets. 
\end{proof}

\section{\revision{Additional Experiments}}
\label{sec:additional_experiments}

\subsection{\revision{Real-world Communication Setup}}

\begin{table}[htb]
\caption{\revision{Running time (seconds) of \epmpd in LAN and WAN settings}}
\label{t:real_world}
\revision{
\resizebox{\columnwidth}{!}{
\begin{tabular}{|c|c|cc|cc|}
\hline
\multirow{2}{*}{\textbf{\begin{tabular}[c]{@{}c@{}}Client\\ Count\end{tabular}}} & \multirow{2}{*}{\textbf{\begin{tabular}[c]{@{}c@{}}Dataset\\ Size\end{tabular}}} & \multicolumn{2}{c|}{\textbf{LAN}}          & \multicolumn{2}{c|}{\textbf{WAN}}          \\ \cline{3-6} 
                                                                                 &                                                                                  & \multicolumn{1}{c|}{\epmpdone} & \epmpdtwo & \multicolumn{1}{c|}{\epmpdone} & \epmpdtwo \\ \hline
30                                                                               & $2^{15}$                                                                         & \multicolumn{1}{c|}{48.24}     & 119.48    & \multicolumn{1}{c|}{49.69}     & 156.41    \\ \hline
50                                                                               & $2^{14}$                                                                         & \multicolumn{1}{c|}{47.73}     & 103.77    & \multicolumn{1}{c|}{80.76}     & 126.8     \\ \hline
100                                                                              & $2^{12}$                                                                         & \multicolumn{1}{c|}{33.92}     & 50.79     & \multicolumn{1}{c|}{49.68}     & 88.08     \\ \hline
150                                                                              & $2^{11}$                                                                         & \multicolumn{1}{c|}{37.09}     & 38.48     & \multicolumn{1}{c|}{44.42}     & 63.85     \\ \hline
\end{tabular}
}
}
\end{table}

\revision{
The experiments in Section~\ref{sec:benchmarks} use an implementation of the \epmpd protocol in the ideal setting to estimate the total computation time and the computation time of the clients and \tee. Data transfer in the ideal setting is free as the various parties directly access the data they share with each other. To estimate the real-world running time, we re-implement the \epmpd protocol using socket programming. Similar to the experiments in Section~\ref{sec:benchmarks}, the clients and \tee run sequentially on the same machine, and the real-world distributed running time is estimated assuming the clients have run simultaneously. We set the network bandwidth and delay using the Linux \texttt{tc} utility and create configurations for the LAN and WAN settings. For the LAN setting, we set the network bandwidth to 1 Gbits/s and the delay to 0.02 ms round trip time (RTT). In the WAN setting, we follow the configuration in \cite{smc_dp} where the network bandwidth is 100 Mbits/s and the delay is 100 ms (RTT). The experiments were conducted on a laptop with an Intel i7-8750H (12) @ 4.100GHz CPU, 16GB RAM, and Ubuntu 22.04.1 LTS.
}

\revision{
Table~\ref{t:real_world} shows the results in the LAN and WAN settings. We fix the duplication percentage to 30\% and vary the number of clients between 30 and 150. The dataset size is varied between $2^{\st 11}$ and $2^{\st 15}$. We observe that in all cases the \epmpd protocols efficiently deduplicate the datasets in under 3 minutes. Additionally, the overheads incurred because of the higher latency in WAN is negligible compared to the LAN. 
}

\subsection{\revision{Increased Client Count}}

\begin{table}[htb]
\caption{\revision{Running time (seconds) of \epmpd, clients, and \tee with large number of clients}}
\label{t:high_clients}
\revision{
\resizebox{\columnwidth}{!}{
\begin{tabular}{|c|cc|cc|cc|}
\hline
\multirow{2}{*}{\textbf{\begin{tabular}[c]{@{}c@{}}Client \\ Count\end{tabular}}} & \multicolumn{2}{c|}{\textbf{Total}}        & \multicolumn{2}{c|}{\textbf{Client}}       & \multicolumn{2}{c|}{\textbf{\tee}}         \\ \cline{2-7} 
                                                                                  & \multicolumn{1}{c|}{\epmpdone} & \epmpdtwo & \multicolumn{1}{c|}{\epmpdone} & \epmpdtwo & \multicolumn{1}{c|}{\epmpdone} & \epmpdtwo \\ \hline
100                                                                               & \multicolumn{1}{c|}{718.84}    & 3822.52   & \multicolumn{1}{c|}{305.04}    & 28.14     & \multicolumn{1}{c|}{1154.38}   & 1076.22   \\ \hline
150                                                                               & \multicolumn{1}{c|}{279.46}    & 1438.31   & \multicolumn{1}{c|}{113.25}    & 7.13      & \multicolumn{1}{c|}{491.97}    & 405.34    \\ \hline
200                                                                               & \multicolumn{1}{c|}{531.22}    & 1915.73   & \multicolumn{1}{c|}{144.72}    & 7.19      & \multicolumn{1}{c|}{1149.73}   & 541.25    \\ \hline
\end{tabular}
}
}
\end{table}

\revision{
Table~\ref{t:high_clients} shows the total running time, client time, and \tee time for scenarios with a high number of clients. The dataset size is fixed to $2^{\st 17}$ for 100 clients and $2^{\st 15}$ for 150 and 200 clients. The duplication level for all experiments is set to $30\%$. The total number of samples in the combined datasets are $\approx13$ million, $\approx4.9$ million, and $\approx6$ million samples in the 100, 150, and 200 client count settings. All experiments were conducted on the same machine and experimental setup as the benchmark experiments in Section~\ref{sec:benchmarks}. From the table, we see that the \epmpd protocols can efficiently scale to large number clients while also ensuring each client's computation is low. Notably, \epmpdone takes $\approx530$ seconds to deduplicate the dataset with $\approx6$ million elements held across 200 clients. With \epmpdtwo, in the same setting, the running time of each client is just 7 seconds.
}

\section{Artifact Appendix}


 

\subsection{Description \& Requirements}




The artifact is hosted online in a public GitHub repository and can be accessed at \url{https://github.com/vdasu/deduplication}. Additionally, the artifact is also available on Zenodo (DOI: \url{https://doi.org/10.5281/zenodo.14251896}).

\subsubsection{How to access}
Simply clone the GitHub or Zenodo repository on your local machine. 


\subsubsection{Hardware dependencies}
The artifact comprises two separate parts. The first implements the \epmpd protocol as a Python library. The second creates an FL setting to analyze the effect that duplicates have on GPU training time and performance. The first part of the artifact can be run on any commodity hardware. The second part requires a GPU system with at least 50GB VRAM. For example, we use a server with the NVIDIA RTX A6000 GPU.

\subsubsection{Software dependencies} 
A detailed description of the software dependencies is provided in the repository in the form of Python \texttt{requirements.txt} and Anaconda \texttt{environment.yml} files. 

\subsubsection{Benchmarks} 
The first part of the artifact that implements the \epmpd protocol uses synthetic data that can be generated by the library itself. The second part of the artifact that implements the FL training requires external datasets. These datasets are provided in a Google Drive link that can be found in the repository.

\subsection{Artifact Installation \& Configuration}

We provide detailed step-by-step instructions to set up the environment, install the artifact, and configure it to run the experiments in the README files in the repository. 




\subsection{Major Claims}



The major claims in the paper are:

\begin{itemize}
    \item (C1): The \epmpd protocol securely removes all pairwise duplicates among the datasets held by two or more clients.
    \item (C2): The total \epmpd running time and \tee running time increase linearly, for both \epmpdone and \epmpdtwo, as the number of clients increases while the dataset size and duplication percentage are fixed. The client running time for \epmpdone increases linearly while the client running time for \epmpdtwo is constant. This claim is supported by Figure~\ref{fig_client}.
    \item (C3): The total \epmpd running time, \tee running time, and client running increases linearly, for both \epmpdone and \epmpdtwo, as the dataset size increases while the number of clients and duplication percentage is fixed. This claim is supported by Figure \ref{fig_dsize} 
    \item (C4): The \epmpd running time, \tee running time, and client running remain constant for \epmpdtwo and decrease linearly for \epmpdone as the duplication percentage increases while the number of clients and dataset set is fixed. This claim is supported by Figure \ref{fig_dp}.
    \item (C5): The overall running time of \epmpdone is less than the running time of \epmpd two. This claim is supported by the \epmpd running times in Figures \ref{fig_client}, \ref{fig_dsize}, and \ref{fig_dp}.
\end{itemize}

Other claims in the paper are:

\begin{itemize}
    \item (C6): The GPU training time and perplexity of the trained models improve as the number of duplicates among the clients decreases. The duplicated setting with no duplicates offers the best performance in both, running time and model utility. This claim is supported by Tables \ref{t:pp} and \ref{t:time}.
\end{itemize}

\subsection{Evaluation}

Experiments E1 and E2 require a common setup. Reading the documentation in the README files in the repository and setting up the codebase requires 10 human minutes. Experiment E1 requires 1 compute minute and Experiment E2 requires 20 compute minutes for a small scale setting. More details about the feasibility of running the large-scale setting used in the paper are provided in the README. Both, the small-scale and large-scale settings, can be used to validate the claims.

\subsubsection{Experiment (E1)} [1 compute-minute]: This experiment validates claim C1.

\textit{[How to]} Experiment E1 simply executes the \epmpd protocol and compares it to a naive non-private deduplication. The \epmpd protocol has been implemented as a Python library in the artifact. The experiment first deduplicates datasets with 4 elements held by 3 clients with \epmpd and naive non-private deduplication. Then, it compares the clients' deduplicated datasets with \epmpd and non-private deduplication.

\textit{[Preparation \& Execution]}
The README inside the \texttt{EP\_MPD} folder in the repository contains an example to perform \epmpd deduplication and compares this to naive non-private deduplication. A code snippet is provided in the ``Usage'' section of the README file creates a setting with 3 clients and datasets with 4 elements each. The ``Installation'' section provides a detailed description of how to set up the environment and install the \epmpd library.

\textit{[Results]}
The outcome of naive non-private deduplication should be the same as \epmpd.

\subsubsection{Experiment E2} [20 compute minutes]: This experiment validates claims C2-C4 on the effect of client count, dataset size, and duplication percentage on the total running time of the \epmpd protocol, clients, and \tee. Specifically, they can be used to generate Figure \ref{fig_client}, \ref{fig_dsize}, and \ref{fig_dp}. These figures are used to infer the running time trends.

\textit{[How To]} Experiment E2 is designed to execute the \epmpd protocol for varying configurations and generate detailed timing information during the protocol execution. The timing information is dumped to log files which are used to generate figures that validate claims C2 - C4. The generated figures need to be manually inspected to validate the claims. For example, the generated figures should have a linear increase in \epmpdone client running time and constant \epmpdtwo client running as we vary the number of clients. This is highlighted in Figure \ref{fig_client:client}.


\textit{[Preparation \& Execution]}
The README inside the \texttt{EP\_MPD} folder in the repository provides a detailed description of how to run the experiment. The ``Reproduce the paper's results'' section of the README provides steps to either generate the paper's figures from log files or re-run the large-scale experiments in the paper. The ``Small Scale Experiments'' section in the README provides steps to run small-scale experiments that can be completed in under 20 minutes on commodity hardware.

\textit{[Results]}
The experiment should run without any errors and produce the log files. The generated figures should follow the same trends observed in Figure \ref{fig_client}, \ref{fig_dsize}, and \ref{fig_dp} and validate claims C2 - C4 on the running time of \epmpd, the clients, and \tee.

Experiment E3 validates claim C6 and runs the FL experiments in the paper. This experiment requires a GPU with 50GB VRAM. Since the experiment deals with the federated learning of LLMs and simulates all clients on a single machine, we cannot create a small-scale version. We note that claim C6 and experiment E3 are used to motivate the need for a deduplication protocol in FL and show the effect of varying duplication percentages on FL performance. Our main contribution is the \epmpd protocol whose claims are validated in the previous experiments.

\subsubsection{Experiment (E3)}
[10 human minutes + 4 compute-days]: This experiment fine-tunes the GPT-2 Medium and Large LMs for 10 clients using FL with various datasets and duplication percentages.

\textit{[How to]} Experiment E3 analyzes the effect of duplicates in training data by manually implanting duplicates, with varying duplication levels, in datasets commonly used to fine-tune LLMs. The experiment fine-tunes LLMs with four duplication levels for each dataset and LLM. Namely, ``Deduplicated'', ``10\%'', ``20\%'', and ``30\%''. The ``Deduplicated'' setting corresponds to the case when a deduplication protocol like \epmpd is used to remove duplicates before training. The experiment generates log files that store the GPU training time and model perplexity. Each combination of duplication level, dataset, and LLM needs to run separately. The paper considers 44 different combinations, and experiment E3 encompasses all of them.

\textit{[Preparation \& Execution]}
The artifact uses an Anaconda environment to manage dependencies and install the required packages. The README inside the \texttt{FL} folder in the repository provides a detailed description of how to run the experiment under the ``Installation Instructions'' and ``Usage'' sections. 

\textit{[Results]}
The generated log files can be used to create Table \ref{t:pp} and \ref{t:time}. The generated log files should highlight the same trends observed in Table \ref{t:pp} and \ref{t:time}. The GPU training time should increase as the number of duplicates increases. Similarly, the model perplexity should worsen (increase) as the number of duplicates increases.

\subsection{Customization}

All experiments can be customized by choosing a different set of parameters from the ones used in the paper. The README files in the repository provide a detailed description of how to run the experiments with different parameters. 





\end{document}